\documentclass[10pt,english,floatfix,nofootinbib,superscriptaddress,aps,prd,preprint]{revtex4}
\usepackage[utf8]{inputenc}       

\usepackage[english]{babel}       
\usepackage[utf8]{inputenc}
\usepackage{textcomp}
\usepackage{amsmath,amsthm,amssymb,amsfonts,mathrsfs,amsbsy} 
\usepackage{tensor}               
\usepackage{slashed}  
\usepackage{bbm}
\usepackage{esint}  
\usepackage{braket}
\usepackage[a4paper, margin=1.2cm]{geometry}
\usepackage{cancel}               
\usepackage{tikz}
\usetikzlibrary{arrows.meta,positioning,calc}
\usepackage{graphicx}             
\usepackage{natbib}
\usepackage{float}                
\usepackage{subfig}               
\usepackage[font=small,labelfont=bf]{caption} 
\usepackage{multirow}             
\usepackage{array}                
\usepackage{tikz}                 
\usetikzlibrary{quotes,angles,arrows,decorations.markings} 
\usepackage{makecell} 

\usepackage{lipsum}

\usepackage{xcolor}
\colorlet{RED}{red} 
\usepackage{color}                
\usepackage{textcomp}             
\usepackage{units}                

\newcommand{\be}{\begin{align}}
\newcommand{\ee}{\end{align}}
\newcommand{\Be}{\begin{eqnarray}}
\newcommand{\Ee}{\end{eqnarray}}

\newcommand{\mincir}{\raise
-3.truept\hbox{\rlap{\hbox{$\sim$}}\raise4.truept\hbox{$<$}\ }}
\newcommand{\magcir}{\raise
-3.truept\hbox{\rlap{\hbox{$\sim$}}\raise4.truept\hbox{$>$}\ }}

\newcolumntype{Y}{>{\centering\arraybackslash}X}

\usepackage{epsfig} 
\usepackage{hyperref}             
\hypersetup{
    colorlinks=true,              
    breaklinks=true,              
    citecolor=blue,               
    linkcolor=[rgb]{0,0.5,0.9},   
    urlcolor=red,                 
    filecolor=green               
}

\newcommand{\ie}{\begin{equation}}
\newcommand{\fe}{\end{equation}}
\newcommand{\se}{\begin{eqnarray}}
\newcommand{\ff}{\end{eqnarray}}

\begin{document}

\title{Entanglement, equivalence principle, and HBAR entropy in a new bumblebee black hole}


\author{A. A. Ara\'{u}jo Filho}
\email{dilto@fisica.ufc.br}
\affiliation{Departamento de Física, Universidade Federal da Paraíba, Caixa Postal 5008, 58051--970, João Pessoa, Paraíba,  Brazil.}
\affiliation{Departamento de Física, Universidade Federal de Campina Grande Caixa Postal 10071, 58429-900 Campina Grande, Paraíba, Brazil.}
\affiliation{Center for Theoretical Physics, Khazar University, 41 Mehseti Street, Baku, AZ-1096, Azerbaijan.}


\author{Wentao Liu}
\email{wentaoliu@hunnu.edu.cn}
\affiliation{Lanzhou Center for Theoretical Physics, Key Laboratory of Theoretical Physics of Gansu Province, 
Key Laboratory of Quantum Theory and Applications of MoE,
Gansu Provincial Research Center for Basic Disciplines of Quantum Physics, 
Lanzhou University, Lanzhou 730000, China}
\affiliation{Institute of Theoretical Physics $\&$ Research Center of Gravitation,
Lanzhou University, Lanzhou 730000, China}


\date{\today}

\begin{abstract}

We investigate quantum {correlations, atomic excitation,} and {horizon brightened acceleration radiation (HBAR) entropy near} a {recently proposed} bumblebee black hole with spacelike or lightlike {Lorentz violating} vacuum {configurations}. Using {localized scalar field packets and} the near horizon Rindler correspondence, we {evaluate} the logarithmic negativity and mutual information {as functions of observer} position, {packet} frequency, and {geometric deformation}. {Configurations with} the {same geometry yield identical correlation measures. Positive geometric deformations} partially {counteract} entanglement degradation, with the largest differences occurring at low frequencies. We {also compute} the excitation {probability} of a freely falling two level atom {using} the {time} coordinate {normalized} at spatial infinity. {At} leading near horizon order and within the {frequency hierarchy} adopted, {its thermal factor agrees} with the {local flat spacetime acceleration radiation result}. Finally, we {derive} the HBAR entropy {flux and its relation} to the {radiative horizon area loss using the canonical conserved energy. The area coefficient depends on the causal character of the vacuum even for configurations with identical geometries: the spacelike case acquires a factor set by the} bumblebee {coupling, while the lightlike case gives the usual one quarter coefficient. These coefficients reproduce} the corresponding {thermodynamic} entropy {factors}.
\end{abstract}

\maketitle

\tableofcontents

\section{Introduction }

Ideas originating in quantum information science have transformed the way correlations and measurements are interpreted in fundamental physics \cite{Hu:2018mni,Boschi:1997dg,Pan:2000vtd}. 
Among these developments, entanglement has taken on a central role as a diagnostic tool for identifying how relativistic motion and spacetime structure influence quantum systems \cite{Aziz:2025ypo}. Long before gravity is introduced, even Minkowski spacetime already reveals this sensitivity: observers undergoing uniform acceleration do not agree with inertial ones on the amount of shared quantum correlations. This mismatch arises because accelerated detectors effectively decompose the field into Rindler modes, leading to a redistribution of correlations between causally disconnected regions. The resulting loss of accessible entanglement, commonly associated with the Unruh effect, provides a clear illustration of how kinematics alone can reshape quantum correlations \cite{Unruh:1976db,Adesso:2007wi,Barros:2023kor,Barros:2024wdv,Kollas:2022wgj,Barros:2025arh,Barros:2025din}.

Once spacetime curvature is taken into account, the fate of quantum correlations changes in a qualitatively similar way: horizons act as filters that redistribute vacuum entanglement into thermalized contributions, altering the correlations accessible to localized observers \cite{Fuentes-Schuller:2004iaz,Fuentes:2010dt,Alsing:2006cj}. 
This recognition has driven a broad program investigating relativistic quantum information in gravitational environments, first within general relativity and later across a variety of modified gravity scenarios \cite{Martin-Martinez:2010yva,Liu:2024wpa,Liu:2015oat,Wu:2023sye,Liu:2023lok,Sen:2023sfb,Li:2022pwa,Wu:2022lmc,Wu:2022xwy,Babakan:2024abb,Wu:2023spa,Wu:2024qhd,AraujoFilho:2025rwr,AraujoFilho:2025hkm,Wu:2025euf,Li:2025jlu,Wu:2024urq,Huang:2024vyc,Liu:2025hcx,Wu:2025ncd,Du:2024ihk,Liu:2026vvp}. 
A fundamental element in this line of research has been the adoption of operational detector-based frameworks. In particular, Unruh--DeWitt detectors provide a remarkable way to track how quantum fields mediate correlations between localized systems moving along relativistic trajectories \cite{Louko:2006zv,Du:2025jiv,Li:2024wtc,Wu:2025qqu}. Through this approach, phenomena such as entanglement harvesting, correlation extraction, and their dependence on spacetime geometry have been quantitatively analyzed, establishing a direct link between field--theoretic structure and observable quantum correlations \cite{Henderson:2017yuv,Tjoa:2020eqh,Cong:2018vqx,Foo:2020dzt,Zhang:2020xvo,Gallock-Yoshimura:2021yok,Bueley:2022ple,Zhou:2021nyv,Li:2025bzd,Liu:2025zts,Tang:2025mtc,Liu:2025bpp}.

{The possibility that Lorentz symmetry may be an effective property of spacetime, rather than an exact principle, has received considerable attention in attempts to connect gravity with quantum phenomena \cite{colladay1997cpt,kostelecky1989spontaneous,kostelecky2004gravity,kostelecky2011data,kostelecky1999constraints}. In this setting, spontaneous symmetry breaking occurs when an additional dynamical field acquires a nonzero vacuum expectation value and selects a preferred spacetime direction. Bumblebee gravity realizes this mechanism through a vector field $B_{\mu}$ governed by a self-interaction potential $V(B_{\mu}B^{\mu}\mp b^{2})$, whose minimum fixes the norm of the vacuum configuration \cite{Bluhm:2023kph,Maluf:2013nva,bluhm2005spontaneous,Bluhm:2019ato,Maluf:2014dpa,bluhm2008spontaneous,Liu:2025lwj,Casana:2017jkc}. The symmetry breaking is therefore generated dynamically rather than imposed explicitly, and the resulting vacuum expectation value modifies the gravitational sector through the preferred direction that it defines \cite{Liu:2024axg,Chen:2025ypx,Liu:2022dcn,Liu:2024oeq,Deng:2025uvp,Tang:2025eew,Sekhmani:2025zen,Li:2025itp,Lai:2025nyo}. Around the vacuum, the modes satisfying the fixed-norm constraint behave as massless gauge--like excitations, whereas those that perturb the potential away from its minimum are massive \cite{bluhm2005spontaneous,bluhm2008spontaneous}. Related realizations of spontaneous Lorentz symmetry breaking based on antisymmetric tensor fields have also been investigated \cite{Yang:2023wtu,Liu:2024oas,Liu:2024lve,Liu:2025fxj,Yu:2025odj,Deng:2025atg,Gu:2025lyz,Guo:2023nkd,Xia:2025hwt,Liu:2026khl,Liu:2026cxs}.}

Extending the bumblebee mechanism to curved spacetime inevitably coupled the vacuum vector configuration to the gravitational field equations, transforming Lorentz violation into a genuinely geometric effect \cite{Bertolami:2005bh}. From this point, several independent research directions emerged. One stream concentrated on strong-gravity systems, where compact objects provided a natural arena to test the consequences of symmetry breaking. The black hole solution proposed in \cite{Casana:2017jkc} quickly became a reference geometry, serving as the starting point for investigations of modified horizon physics. Within this setting, analyses revealed departures from standard behavior in quantum correlations near the horizon \cite{Liu:2024wpa} as well as changes in particle creation mechanisms and radiative processes \cite{AraujoFilho:2025hkm}.  Parallel efforts reformulated the symmetry-breaking scheme using antisymmetric tensor fields, leading to Lorentz--violating black hole solutions in the Kalb-Ramond framework and enlarging the class of admissible geometries \cite{AraujoFilho:2024ctw}. A separate line of development addressed large–scale and astrophysical phenomena. The same geometric modifications were also shown to affect the behavior of gravitational waves, yielding propagation characteristics that depart from those predicted by general relativity \cite{Liang:2022hxd,amarilo2024gravitational}.  Further extensions altered the gravitational sector more directly by introducing a cosmological constant into the bumblebee setup, allowing the analysis of modified vacuum structures and related properties in these generalized backgrounds \cite{Maluf:2020kgf,Uniyal:2022xnq}.

Since its original formulation on a simple static background \cite{Casana:2017jkc}, the bumblebee scenario has undergone a substantial diversification. One of the most significant shifts occurred when the construction was embedded in the \textit{metric--affine} formalism, where the metric and connection are treated as independent variables and additional geometric structure naturally arises. Within this perspective, a static Lorentz--violating solution was first obtained in \cite{Filho:2022yrk}, and this result later served as the foundation for an axially symmetric, rotating configuration \cite{AraujoFilho:2024ykw}. These geometries subsequently motivated further generalizations, including non--commutative deformations of the bumblebee framework \cite{AraujoFilho:2025rvn} and parallel developments based on antisymmetric tensor fields, such as those appearing in Kalb--Ramond gravity \cite{AraujoFilho:2025jcu}.

The scope of these investigations has also moved beyond black-hole spacetimes. It was shown that a vector field constrained to a fixed norm can act as a source capable of supporting wormhole solutions or modifying their traversability conditions \cite{Magalhaes:2025nql,AraujoFilho:2024iox,Ovgun:2018xys,Magalhaes:2025lti}. In a related but distinct direction, black-bounce geometries sustained by $\kappa$--essence fields were proposed while still retaining the hallmark features of spontaneous Lorentz symmetry breaking \cite{Pereira:2025xnw}. Propagation effects constitute another central theme. The bending of neutrino trajectories has been explored in several realizations of Lorentz--violating gravity, including purely \textit{metric} bumblebee models \cite{Shi:2025plr}, \textit{metric--affine} extensions \cite{Shi:2025ywa}, and tensorial generalizations of the symmetry-breaking sector \cite{Shi:2025rfq,Shi:2025xkd}. Beyond lensing, additional studies have addressed neutrino phenomenology more broadly, encompassing constraints, oscillation-related features, and other observational consequences within bumblebee gravity \cite{Khodadi:2021owg,Khodadi:2023yiw,Khodadi:2022mzt,Khodadi:2022dff}.

Furthermore, the landscape of Lorentz--violating black holes has recently broadened with the construction of novel geometries that arise explicitly from the bumblebee framework rather than from phenomenological deformations \cite{Zhu:2025fiy,Liu:2025oho}. One such static configuration was subsequently examined in depth in \cite{AraujoFilho:2025zaj}, where its geometric structure, dynamical properties, and observational constraints were systematically explored. The same background was later employed to study particle propagation effects, with particular emphasis on neutrino oscillations \cite{Shi:2025tvu}. Progress has not been limited to nonrotating cases. Starting from the static solution as a seed, an axially symmetric extension was generated through a refined version of the Newman--Janis algorithm, yielding a rotating bumblebee black hole geometry \cite{Kumar:2025bim}. This rotating spacetime has already motivated further investigations, including analyses of matter accretion processes \cite{Shi:2025hfe} and studies of quantum emission and radiative phenomena associated with the horizon \cite{Heidari:2025oop}.

The present work {examines} the near horizon {quantum physics} of {this recently proposed} bumblebee {geometry} through {three} complementary probes. {Initially}, {we evaluate} the logarithmic negativity and mutual information {of localized scalar field packets shared by a freely falling observer} and {a static observer}. The correlation measures {coincide for spacelike} and {lightlike} vacuum configurations with the {same geometry, while positive geometric deformations partially counteract their degradation. Moreover, we calculate} the {excitation probability} of a freely falling two level atom {using the asymptotically normalized time coordinate. At leading} near {horizon order and within} the {adopted frequency hierarchy}, {its thermal factor agrees} with the corresponding flat spacetime {acceleration radiation} result. Finally, {we derive} the {HBAR} entropy {flux and relate it} to the {\hypersetup{citecolor=red,linkcolor=red}radiative horizon area loss through the canonical conserved energy. This relation distinguishes the two vacuum configurations even when their geometries coincide. Section~\ref{sec:background} presents the adopted solution, the quantum field and detector framework}, and the {\hypersetup{citecolor=red,linkcolor=red}thermodynamic inputs from the literature. Section~\ref{sec:results} develops the correlation,} atomic {response}, {and HBAR results, and Sec}.{\hypersetup{citecolor=red,linkcolor=red}~\ref{Sec:Conclusion} summarizes our findings.}


{\section{{Background and Formalism}}\label{sec:background}}

{\hypersetup{citecolor=red,linkcolor=red}This section collects the black hole geometry and thermodynamic quantities adopted from Refs.~\cite{Liu:2025oho,Zhu:2025fiy}, the standard Rindler and quantum correlation framework~\cite{Fuentes-Schuller:2004iaz,Martin-Martinez:2010yva,Bruschi:2010yva}, and the atomic and HBAR methods~\cite{scully2018quantum,sen2022equivalence}. Intermediate coordinate transformations and detector kinematics are written explicitly for the adopted metric. The applications yielding the parameter dependent correlation measures, atomic response, and HBAR area relation are developed in Sec.~\ref{sec:results}.}


{\subsection{{Adopted bumblebee black hole geometry}}\label{sec:geometry}}

{We adopt} the {static black hole solution obtained in the literature for spacelike and lightlike} vacuum expectation values (VEVs) \cite{Liu:2025oho,Zhu:2025fiy}, whose static and spherically symmetric geometry is characterized by the metric
\begin{equation}\label{ds2}
\mathrm{d} s^2 = -\frac{1}{1+\alpha\ell}\left(1-\frac{2M}{r} \right)\mathrm{d}t^2 + \frac{1+\alpha\ell}{(1-\tfrac{2M}{r})}\mathrm{d}r^2 + r^2 \mathrm{d}\Omega^{2},
\end{equation}
{where $M$ is the metric mass parameter, $r$ is the areal radius, $\mathrm{d}\Omega^2$ is the line element on the unit two sphere, and} $\ell=\Tilde{\xi}b^2$ {is the dimensionless bumblebee coupling}. {Here, $\Tilde{\xi}$ denotes the nonminimal coupling between the bumblebee field and curvature, while $b>0$ is the constant amplitude scale used to parameterize the vacuum expectation value. In the notation of Ref.~\cite{Zhu:2025fiy}, the vacuum condition reads $b_\mu b^\mu+s b^2=0$, where $s$ fixes the causal character of the vacuum; in particular, the invariant norm vanishes for the lightlike {configuration} even though $b$ remains the amplitude scale entering the component parameterization.  } The spacetime geometry depends only on the combination $\alpha \ell$, which has been denoted as $\chi=\alpha \ell$ in some works to characterize the overall effect \cite{Shi:2025xkd}, whereas the nature of the vacuum expectation value is encoded in the choice of $\alpha$. {Here, $\alpha$ is the dimensionless coefficient specifying the vacuum parametrization.} Following {\hypersetup{citecolor=red,linkcolor=red}Ref.~\cite{Zhu:2025fiy}}, {its} spacelike and lightlike {values are}
\begin{equation}
\text{\rm Spacelike}: \alpha=\beta^2+1, \quad \text{\rm Lightlike}: \alpha=\beta^2,
\end{equation}
with $\beta\in \mathbb{R}$. For each fixed value of $\alpha\ell$, suitable choices of $\beta$ generate spacelike and lightlike VEVs that share the same metric structure but differ in the underlying Lorentz--violating vacuum configuration. In what follows, we refer to $\ell$ as the Bumblebee parameter, which quantifies the strength of Lorentz symmetry breaking, and to $\beta$ as the LZ parameter, which characterizes the vacuum orientation of the vector field (i.e., the relative weight of its temporal and radial components). For the spacelike {configuration}, the limit $\beta=0$ corresponds to a purely spacelike vacuum configuration, and the solution reduces to the standard Bumblebee black hole.  In contrast, for the lightlike {configuration}, the limit $\beta=0$ eliminates the background vector field and the solution exactly reduces to the Schwarzschild spacetime of general relativity.

{For reference, we recall} the {standard acceleration and} surface gravity {construction for} the {adopted metric}. For a static observer, the four--velocity takes the form
\begin{align}
u^\mu=\left\{u^0,0,0,0\right\},
\end{align}
where $u^0$ is fixed by the normalization condition $u^\mu u_\mu = -1$.  The corresponding four-acceleration is defined as
\begin{align}\label{amu}
a^\nu=u^\mu\nabla_\mu u^\nu=u^\mu\partial_\mu u^\nu+\Gamma^\nu_{\mu\rho}u^\mu u^\rho.
\end{align}
For the static spacetime described by Eq.~(\ref{ds2}), the surface gravity can be extracted from the near--horizon expansion of the lapse function, yielding \cite{Heidari:2025oop}
\begin{equation}
\begin{aligned}
\kappa=\frac{1}{4M(1+\alpha\ell)}.
\end{aligned}
\end{equation}
{Here $\kappa\equiv\kappa_t$ refers to the Killing field $\partial_t$.} In semiclassical gravity, the {associated} Hawking temperature is given by $T \!=\! \frac{\kappa}{2\pi}$ as discussed in Ref. \cite{Wald:1993nt}.


{We use $\chi=\alpha\ell$ for the geometric deformation and $q=1+\chi>0$ for the radial metric coefficient. The parameter $\beta$ introduced above fixes the relative temporal and radial components of the vacuum vector. These definitions apply throughout the formalism and results below.}


{\subsection{{Rindler representation and field modes}}\label{sec:rindler}}

{\hypersetup{citecolor=red,linkcolor=red}The flat spacetime Rindler construction is standard~\cite{Unruh:1976db,sen2022equivalence}.} We begin with a flat spacetime background restricted to two dimensions, where one temporal and one spatial coordinate define the geometry. {Using the signature adopted {here}, the line element reads}
\begin{equation}\label{gghhhhh}
\mathrm{d}s^2 = {-}\mathrm{d}t^2
{+} \mathrm{d}z^2\,.
\end{equation}
{We now consider} worldlines characterized by uniform proper acceleration. {Their coordinates are parametrized by the proper time $\tau$ as}
\ie
t(\tau)=\frac{1}{a} \sinh\left({a\tau}\right)
\label{sdddd},
\fe
\ie
z(\tau)=\frac{1}{a}
\cosh\left({a\tau}\right)\,.
\label{assxxx}
\fe
{We introduce the Rindler coordinates by applying the following transformations} to Eq.~(\ref{gghhhhh}):
\begin{align}
t&=\rho\sinh\left({\tilde{a}\tilde{t}}\right),
\label{dssssss}\\
z&=\rho\cosh\left({\tilde{a}\tilde{t}}\right)\,.
\label{aaaaasss}
\end{align}

Inserting the transformations (\ref{dssssss})--(\ref{aaaaasss}) into Eq.~(\ref{gghhhhh}) yields {the Rindler line element}
\ie
\label{sdfffff}
\mathrm{d}s^2 = {-} \left(\tilde{a}\rho\right)^2\mathrm{d}\tilde{t}^2 {+} \mathrm{d}\rho^2\,.
\fe
{For a trajectory at fixed $\rho$, comparison of Eqs.~(\ref{sdddd})--(\ref{assxxx}) with Eqs.~(\ref{dssssss})--(\ref{aaaaasss}) gives $\rho=1/a$ and $a\tau=\tilde{a}\tilde{t}$. Therefore,}
\begin{equation}
\tau=\frac{\tilde{a}\tilde{t}}{a}\,.
\end{equation}
{The corresponding proper acceleration is}
\begin{equation}
\label{rindlerflatacc}
a=\frac{1}{\rho}\,.
\end{equation}
{Here, $\tilde{a}$ fixes the normalization of the Rindler time and is not the proper acceleration of every trajectory at fixed $\rho$.}

{We follow Refs.~\cite{Martin-Martinez:2010yva,Sen:2023sfb} and use the correspondence between the near--horizon region of a static black hole and a Rindler wedge.} {We retain the time coordinate $t$ appearing in Eq.~(\ref{ds2}), without rescaling it at spatial infinity. For convenience, we define}
\begin{equation}\label{metric_functions_ent}
{
q=1+\alpha\ell, \qquad f(r)=1-\frac{2M}{r}, \qquad F(r)=\frac{f(r)}{q}. }
\end{equation}
{The radial sector of the metric is then}
\begin{equation}\label{radial_metric_ent}
{ \mathrm{d}s_{(2)}^2=-F(r)\mathrm{d}t^2+\frac{\mathrm{d}r^2}{F(r)}. }
\end{equation}

For this static bumblebee black hole metric (\ref{ds2}), we begin by examining the behavior of the metric function in the vicinity of the event horizon. Employing the near--horizon approximation, the metric function is expanded as
\begin{equation}
{ F(r)=F'(r_h)(r-r_h)+\mathcal{O}\!\left((r-r_h)^2\right), \qquad F'(r_h)=\frac{2M}{q r_h^2}, \qquad r_h=2M. }
\end{equation}
{The zeroth-order term vanishes because $F(r_h)=0$.}

{For the field calculation, we retain the $(1+1)$-dimensional radial sector of the spacetime \cite{Fuentes-Schuller:2004iaz}. We define the proper distance from the horizon, to leading order, as}
\begin{equation}\label{drdz}
{ \xi=2\sqrt{\frac{r-r_h}{F'(r_h)}}.}
\end{equation}
{Substitution into Eq.~(\ref{radial_metric_ent}) gives}
\begin{align}\label{ds17}
\mathrm{d}s_{{(2)}}^2={-\kappa_t^2\xi^2\mathrm{d}t^2}+\mathrm{d}\xi^2 {+\mathcal{O}(\xi^4)},
\end{align}
{where the surface gravity associated with the unnormalized Killing field $\partial_t$ is}
\begin{equation}\label{kappa_ent}
{ \kappa_t=\frac{F'(r_h)}{2} =\frac{1}{4M(1+\alpha\ell)}, \qquad T_t=\frac{\kappa_t}{2\pi}. }
\end{equation}
{Here $T_t$ is the Hawking temperature defined with respect to the coordinate $t$. Since $g_{tt}\rightarrow-1/q$ at spatial infinity, neither $\kappa_t$ nor $T_t$ is normalized at infinity. Notice that this construction gives the Rindler form directly and does not require an arbitrary integration constant.}

We now introduce a static observer located at a fixed radial position $r_0$, whose proper time is denoted by $\tau$. Since the observer remains at rest with respect to the spatial coordinates, we have $\mathrm{d}r_0=0$, and the line element reduces to
\begin{equation}
-\mathrm{d}\tau^2=-F(r)|_{r=r_0}\mathrm{d}t^2+F(r)^{-1}|_{r=r_0}\mathrm{d}r_0^2=-F(r_0)\mathrm{d}t^2,
\end{equation}
from which the relation between coordinate time and proper time follows immediately as
\begin{equation}
\frac{\mathrm{d}t}{\mathrm{d}\tau}=\frac{1}{\sqrt{F_0}},\quad\quad t=\frac{1}{\sqrt{F_0}}\tau,
\end{equation}
where $F_0 \equiv F(r_0)${, with $F_0=f_0/q$ and $f_0=1-2M/r_0$}.
Substituting this relation into Eq.~(\ref{ds17}), we rewrite the near--horizon metric in terms of the proper time $\tau$ as
\begin{align}\label{dstau}
\mathrm{d}s_{{(2)}}^2=-\frac{{\kappa_t^2}}{F_0}\xi^2 \mathrm{d}\tau^2+\mathrm{d}\xi^2.
\end{align}

{This is the Rindler metric,}
\begin{equation}
\mathrm{d}s^2 = -a_{{\rm R}}^2\xi^2 \mathrm{d}\tau^2 + \mathrm{d}\xi^2,
\end{equation}
from which the effective acceleration parameter is identified as
\begin{equation}\label{aR_ent}
a_{{\rm R}} = \frac{{\kappa_t}}{\sqrt{F_0}}
=\frac{1}{4M \sqrt{(1+\alpha\ell)(1-\frac{2M}{r_0})}}.
\end{equation}
To make the physical interpretation of this parameter explicit, we next evaluate the proper acceleration of a static observer at $r=r_0$ directly in the original black hole spacetime.

For a stationary observer located at an arbitrary fixed position $r$, the proper acceleration is defined as $a=\sqrt{a_\mu a^\mu}$, where the four--acceleration $a^\mu$ is given by Eq.~(\ref{amu}). Evaluating the components of $a^\mu$ and $a_{\mu}$ for the metric (\ref{ds2}), we obtain
\begin{align}\label{admu}
a^\mu=\left\{0,\frac{M}{r^2(1+\alpha\ell)},0,0 \right\}, \quad \quad a_\mu=\left\{0,\frac{M}{r(r-2M)},0,0 \right\},
\end{align}
and the corresponding proper acceleration reads
\begin{align}\label{aaaa}
{a_{\rm prop}(r)}=\sqrt{a^\mu a_\mu}=\frac{M}{r^2}\frac{1}{\sqrt{(1+\alpha\ell)\left(1-\frac{2M}{r} \right)}}.
\end{align}
{The exact proper acceleration in Eq.~(\ref{aaaa}) and the Rindler acceleration in Eq.~(\ref{aR_ent}) are related by}
\begin{equation}
{ \frac{a_{\rm prop}(r_0)}{a_{\rm R}(r_0)}=\frac{4M^2}{r_0^2}.}
\end{equation}
{They coincide only in the near--horizon limit $r_0\rightarrow r_h$. Here $a_{\rm R}$ is the acceleration parameter entering the local Rindler description, whereas $a_{\rm prop}$ is the exact invariant acceleration of the static observer.}

{We next write the near--horizon metric in Kruskal coordinates. The required tortoise coordinate is}
\begin{equation}
r_*=(1+\alpha\ell)\int \frac{1}{(1-\frac{2M}{r})}\mathrm{d}r =(1+\alpha \ell)\left[r+2M\ln\left(\frac{r}{2M}-1\right) \right],
\end{equation}
and define the null coordinates $u=t-r_*$ and $v=t+r_*$. In terms of these variables, the radial sector of the line element (\ref{ds2}) takes the form $\mathrm{d}s_{{(2)}}^2=-F(r)\,\mathrm{d}u\,\mathrm{d}v$. We then introduce the generalized Kruskal coordinates
\begin{equation}
\mathcal{U} = -\frac{1}{{\kappa_t}}e^{-{\kappa_t}(t-r_*)},\quad \mathcal{V}=\frac{1}{{\kappa_t}}e^{{\kappa_t}(t+r_*)},
\end{equation}
which provide a regular coordinate covering across the horizon.
{Positive frequency with respect to the Kruskal time defines the Hartle--Hawking modes. Positive frequency with respect to $t$ in the exterior and interior regions defines the Boulware and partner Boulware modes, denoted by B and $\bar{\mathrm B}$, respectively. At the level of the vacuum states, the standard correspondence may be summarized as}
\begin{equation}
\begin{aligned}\label{ABBt}
\ket{0}_\text{A}\leftrightarrow\ket{0}_\text{M}\leftrightarrow\ket{0}_\text{H},\quad\quad
\ket{0}_\text{R}\leftrightarrow\ket{0}_\text{I~}\leftrightarrow\ket{0}_\text{B},\quad\quad
\ket{0}_{\bar{\text{R}}}\leftrightarrow\ket{0}_\text{IV}\leftrightarrow\ket{0}_{\bar{\text{B}}}.
\end{aligned}
\end{equation}
{The arrows in Eq.~(\ref{ABBt}) refer to this vacuum correspondence. They do not give a one-to-one transformation between arbitrary monochromatic Minkowski and Rindler modes. The latter transformation is Fourier-like, and the single-mode approximation is not valid for a general state {\cite{Bruschi:2010yva,MartinMartinez:2011ib}}.}

{We therefore describe the excitations by normalized wave packets. Let $h_i(\omega)$ be sharply peaked at the local frequency $\omega_i$ measured by Bob. The exterior packet operator is}
\begin{equation}\label{packet_operator_ent}
{ \hat b_i=\int_0^\infty \mathrm{d}\omega\,h_i(\omega)\hat b_\omega, \qquad \int_0^\infty \mathrm{d}\omega\,|h_i(\omega)|^2=1. }
\end{equation}
{For packets satisfying the peaking conditions discussed in Ref.~\cite{Bruschi:2010yva}, the Bogoliubov transformation can be evaluated at the central frequency. Within this narrow--band approximation, the Hartle--Hawking vacuum is written in terms of Boulware packet modes as}
\begin{align}\label{sk0}
\ket{0}^{i}_\text{H}{\simeq}\frac{1}{\cosh\sigma_{i}}
\sum_{{n=0}}^{\infty} \tanh^n\sigma_{i}\ket{n}^{i}_\text{B}
\ket{n}^{i}_{\bar{\text{B}}},
\end{align}
where $\ket{0}_\text{H}=\otimes_j\ket{0}_\text{H}^{j}$ denotes the full Hartle--Hawking vacuum constructed {packet by packet}, and the squeezing parameter is given by
\begin{equation}
\begin{aligned}\label{tansi}
\tanh\sigma_{i}=&\exp{\left(- \frac{\pi\omega_i}{{a_{\rm R}}}\right)} = \exp\left({-4\pi M \omega_i\sqrt{(1+\alpha\ell)\left(1-\frac{2M}{r_0}\right)}} \right).
\end{aligned}
\end{equation}
{Here $\omega_i$ is the packet frequency measured in Bob's local frame. If $\widetilde{\omega}_i$ denotes the Killing frequency conjugate to the unnormalized coordinate $t$, the two frequencies satisfy $\widetilde{\omega}_i=\sqrt{F_0}\,\omega_i$. In other words, Eq.~(\ref{tansi}) can also be written as}
\begin{equation}\label{tansi_coordinate_ent}
{ \tanh\sigma_i =\exp\left(-\frac{\pi\widetilde{\omega}_i}{\kappa_t}\right) =\exp\left[-4\pi M(1+\alpha\ell)\widetilde{\omega}_i\right]. }
\end{equation}
{The explicit dependence on $r_0$ in Eq.~(\ref{tansi}) arises because the local frequency $\omega_i$ is held fixed. If the Killing frequency $\widetilde{\omega}_i$ is held fixed instead, the same relation takes the form of Eq.~(\ref{tansi_coordinate_ent}) and has no explicit dependence on $r_0$.}
{Although the coordinate surface gravity $\kappa_t$ is proportional to $(1+\alpha\ell)^{-1}$, the local acceleration $a_{\rm R}=\kappa_t/\sqrt{F_0}$ is proportional to $(1+\alpha\ell)^{-1/2}$. A constant rescaling of the time coordinate rescales $\kappa_t$ and $\widetilde{\omega}_i$ by the same factor. Their ratio, the squeezing parameter, and the correlation measures expressed in terms of Bob's local frequency are therefore unchanged by this coordinate choice.}
The one-particle Hartle--Hawking state is generated by acting with a creation operator on the vacuum and can be written in the Boulware basis as
\begin{equation}\label{sk1}
\ket{1}_\text{H}^{i} {\simeq} \frac{1}{\cosh^2\sigma_{i}} \sum_{n=0}^{\infty} \tanh^n\sigma_{i} \sqrt{n+1} \ket{n+1}_\text{B}^{i} \ket{n}_{\bar{\text{B}}}^{i}.
\end{equation}
{Eqs.~(\ref{sk0}) and (\ref{sk1}) are leading--order wave--packet relations. For a broad packet, the complete Bogoliubov kernel must be retained.}


For the {atomic and thermodynamic calculations,} the Killing {time is normalized} at spatial infinity. {If $\omega$ is the frequency conjugate to $t$}, we introduce
{
\begin{equation}
\label{normalized_coordinates}\bar{t}=\frac{t}{\sqrt{q}}, \qquad \omega_\infty=\sqrt{q}\,\omega, \qquad \bar{r}_*=\frac{r_*}{\sqrt{q}} =\sqrt{q}\int\frac{\mathrm{d}r}{f(r)},
\end{equation}}
{so that the radial sector of the metric and the corresponding surface gravity become}
{
\begin{equation}
\label{normalized_metric}\mathrm{d}s^2_{(2)} =-f(r)\mathrm{d}\bar{t}^{\,2} +\frac{q}{f(r)}\mathrm{d}r^2, \qquad g_{\bar{t}\bar{t}}\longrightarrow-1, \qquad \bar{\kappa} =\sqrt{q}\,\kappa_t =\frac{f'(r_h)}{2\sqrt{q}} =\frac{1}{4M\sqrt{q}}.
\end{equation}}
{Here} $\omega_\infty$ is the {scalar mode} frequency measured {by} a static observer at infinity{, $\bar r_*$ is the normalized tortoise coordinate, and $\bar\kappa$ is the normalized surface gravity}. {\hypersetup{citecolor=red,linkcolor=red}The corresponding temperature is $T_H=\bar\kappa/(2\pi)$ in units with $k_B=1$.

The flat spacetime transformation, the near horizon Rindler approximation, and the thermal vacuum expansion are established ingredients~\cite{Unruh:1976db,Fuentes-Schuller:2004iaz,Martin-Martinez:2010yva,Sen:2023sfb}. The formulas above express them in the coordinates and parameters of the adopted bumblebee solution. A freely falling atom has vanishing proper four acceleration; the local comparison in Sec.~\ref{sec:atomic_results} concerns its transition probability and thermal factor.}


{\subsection{{Field states and correlation measures}}\label{sec:correlation_formalism}}

{\hypersetup{citecolor=red,linkcolor=red}We use the established scalar field correlation framework~\cite{Fuentes-Schuller:2004iaz,Martin-Martinez:2010yva,Bruschi:2010yva}. The Bell state, partial trace, logarithmic negativity, and mutual information below provide the common formulas evaluated for the bumblebee geometry in Sec.~\ref{sec:correlation_results}.}

\begin{figure}[h]
\centering
\includegraphics[width=0.5\linewidth]{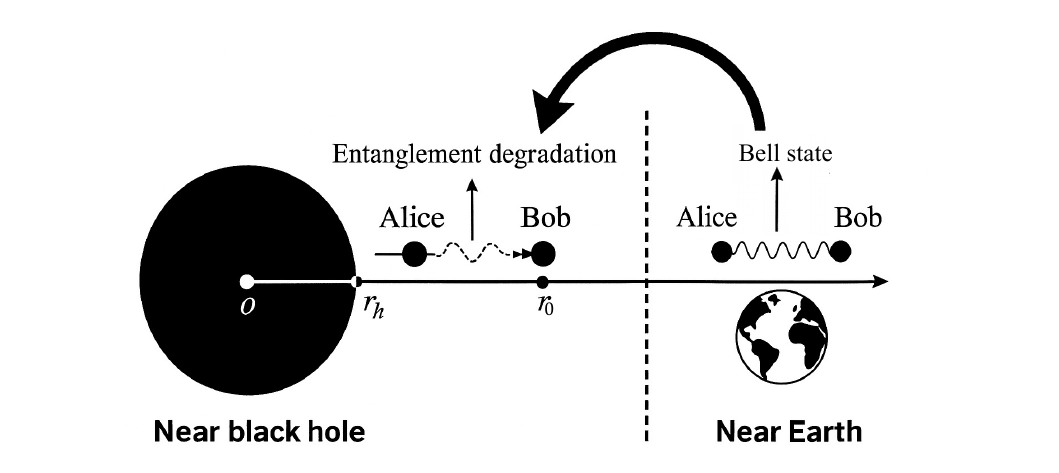}
\caption{Alice and Bob are initially prepared in a {Bell-like state of two localized wave packets}. {Alice follows a freely falling trajectory, whereas Bob is held static at $r_0>r_h$.}}
\label{fig1}
\end{figure}
{The configuration is shown in Fig.~\ref{fig1}. Alice and Bob initially share two localized field packets in a region where the gravitational field is weak. They are then taken toward the black hole. Alice remains in free fall, while Bob is held at a fixed radius outside the horizon. The interior partner of Bob's packet is inaccessible to an exterior observer. After this degree of freedom is traced out, the Alice--Bob state is mixed and its entanglement is reduced.}

The initial bipartite state shared by Alice and Bob is taken to be a maximally entangled Bell state of two {localized field packets}.
{In the wave-packet notation adopted above, it is written as}
\begin{align}\label{simplemax}
\ket{\psi}_\text{AB} = \frac{1}{\sqrt{2}}\left(\ket{0}_{\text{A}}\ket{0}_{\text{H}}^{i} + \ket{1}_{\text{A}}\ket{1}_{\text{H}}^{i}\right),
\end{align}
where Alice (A) freely falls into the black hole and Bob (B) remains static at radius $r_0$ outside the horizon. {The occupation numbers in Eq.~(\ref{simplemax}) refer to localized packets and not to exactly monochromatic Minkowski modes.}

{Substituting Eqs.~(\ref{sk0}) and (\ref{sk1}) into Eq.~(\ref{simplemax}), we obtain the tripartite pure state}
\begin{equation}\label{tripartite_ket_ent}
{ \ket{\psi}_{\mathrm{AB}\bar{\mathrm B}} =\frac{1}{\sqrt{2}\cosh\sigma_i} \sum_{n=0}^{\infty}\tanh^n\!\sigma_i \left[ \ket{0,n,n} +\frac{\sqrt{n+1}}{\cosh\sigma_i}\ket{1,n+1,n} \right]. }
\end{equation}
{The global density operator is $\rho_{\mathrm{AB}\bar{\mathrm B}}=\ket{\psi}\!\bra{\psi}$. Tracing over the inaccessible interior packet gives}
\begin{align}\label{rho_AB_ent}
{\rho_\text{AB}}=&{\frac{1}{2\cosh^2\sigma_i} \sum_{n=0}^{\infty}\tanh^{2n}\!\sigma_i\bigg[ \ket{0,n}\!\bra{0,n} +\frac{\sqrt{n+1}}{\cosh\sigma_i} \left(\ket{0,n}\!\bra{1,n+1}+\ket{1,n+1}\!\bra{0,n}\right) }\nonumber\\ &{\hspace{42mm} +\frac{n+1}{\cosh^2\sigma_i}\ket{1,n+1}\!\bra{1,n+1} \bigg]. }
\end{align}
{This state has unit trace. Alice's reduced state remains maximally mixed, $\rho_{\rm A}=\mathbb{I}_2/2$.}

The logarithmic negativity $\mathcal{N}$ is an entanglement monotone {that provides an upper bound on distillable entanglement}. For a bipartite state $\rho_{\text{AB}}$, it is defined as $\mathcal{N}(\rho_{\text{AB}})=\log_2 \left\| \rho_\text{AB}^{T_\text{A}} \right\|_1$ \cite{Vidal:2002zz,Plenio:2005cwa,Diaz:2023jrf,Xu:2024eqg}, where $T_\text{A}$ denotes the partial transpose with respect to subsystem A and $\|\cdot\|_1=\mathrm{Tr}\sqrt{\rho^\dagger\rho}$ is the trace norm. In the present work, we apply this measure to the reduced density matrix $\rho_\text{AB}$ describing the Alice-Bob subsystem. {The standard partial transpose spectrum gives} the {\hypersetup{citecolor=red,linkcolor=red}following expression~\cite{Fuentes-Schuller:2004iaz,Martin-Martinez:2010yva};} the {dependence on} the {\hypersetup{citecolor=red,linkcolor=red}present geometry enters through $\sigma_i$ in Eq.~(\ref{tansi}):}
\begin{equation}\label{NS1}
\mathcal{N}\left(\rho_\text{AB}\right)\!=\!\log_2||\rho^\text{$\text{T}_\text{A}$}_\text{AB}||_1
\!=\!\log_2\left[\frac{1}{2\cosh^2\!\sigma_{i}}\!+\!\sum^\infty_{n=0}\frac{\tanh^{2n}\sigma_{i}}{2\cosh^2\sigma_{i}}
\sqrt{\Big(\tanh^2\!\sigma_{i}\!+\!\frac{n}{\sinh^2\!\sigma_{i}}\Big)^2\!+\!\frac{4}{\cosh^2\!\sigma_{i}}} \right].
\end{equation}
{The value at $\sigma_i=0$ is understood by continuity.}

The mutual information provides an operationally meaningful quantifier of total correlations, including both classical and quantum contributions. For a density matrix $\rho_\text{AB}$, it is defined as \cite{Fuentes-Schuller:2004iaz}
\begin{align}
I(\rho_\text{AB})=S(\rho_\text{A})+S(\rho_\text{B})-S(\rho_\text{AB}),
\end{align}
where $ S(\rho)=-\text{Tr}(\rho \log_2\rho) =-\sum_i\lambda_{i}\log_2(\lambda_{i})$ is the von Neumann entropy of the density matrix $ \rho $, with $\lambda_{i}$ being its eigenvalues. For {the} scalar {Bell state}, the {standard} entropy {\hypersetup{citecolor=red,linkcolor=red}expressions are~\cite{Fuentes-Schuller:2004iaz,Martin-Martinez:2010yva}}
\begin{equation}
\begin{aligned}
S(\rho_\text{AB})=&-\sum_{n=0}^{\infty}\frac{\tanh^{2n}\sigma_{i}}{2\cosh^2\sigma_{i}} \left(1+\frac{n+1}{\cosh^2\sigma_{i}}\right) \log_2\!\left[\frac{\tanh^{2n}\sigma_{i}}{2\cosh^2\sigma_{i}} \!\left(1+\frac{n+1}{\cosh^2\sigma_{i}}\right)\right].
\end{aligned}
\end{equation}
The reduced density matrix $\rho_\text{B}$ is obtained by tracing out Alice’s degrees of freedom, and its entropy reads
\begin{equation}
\begin{aligned}
S(\rho_\text{B})=&-\sum_{n=0}^{\infty}\frac{\tanh^{2n}\sigma_{i}}{2\cosh^2\sigma_{i}} \left(1+\frac{n}{\sinh^2\sigma_{i}}\right) \log_2\!\left[\frac{\tanh^{2n}\sigma_{i}}{2\cosh^2\sigma_{i}} \!\left(1+\frac{n}{\sinh^2\sigma_{i}}\right)\right].
\end{aligned}
\end{equation}
Since Alice’s reduced state remains maximally mixed, $S(\rho_\text{A})=1$, the mutual information reduces to
\begin{equation}\label{IS1}
\begin{aligned}
I(\rho_\text{AB})=&1-\frac{1}{2}\log_2\tanh^2\sigma_{i} -\sum_{n=0}^{\infty}\frac{\tanh^{2n}\sigma_{i}}{2\cosh^2\sigma_{i}} \left(1+\frac{n}{\sinh^2\sigma_{i}}\right) \log_2\!\left(1+\frac{n}{\sinh^2\sigma_{i}}\right)\\ &+\sum_{n=0}^{\infty}\frac{\tanh^{2n}\sigma_{i}}{2\cosh^2\sigma_{i}} \left(1+\frac{n+1}{\cosh^2\sigma_{i}}\right) \log_2\!\left(1+\frac{n+1}{\cosh^2\sigma_{i}}\right).
\end{aligned}
\end{equation}


{\subsection{{Freely falling atomic trajectory and interaction}}\label{sec:atomic_formalism}}

{\hypersetup{citecolor=red,linkcolor=red}We follow the freely falling two level atom construction of Refs.~\cite{scully2018quantum,sen2022equivalence}. The atomic ground and excited states are denoted by $g$ and $e$, and $\tau$ is the proper time along the geodesic. The trajectory and outgoing scalar modes below specialize this standard detector setup to the adopted metric.}

For radial infall from rest at spatial infinity, the conserved Killing energy {associated with $\partial_{\bar{t}}$} is $\mathcal{E}={1}$. Accordingly,
\begin{equation}
\begin{split}
\tau(r) &=-\sqrt{q}\int\frac{\mathrm{d}r}{\sqrt{1-f(r)}} \\ &\approx-\frac{1}{3} \left(\sqrt{2 \,q}\,M\left(\frac{r}{M}\right)^{3/2}\right) +C+\mathcal{O}(\chi^2).
\end{split}
\end{equation}
Here, the factor $\sqrt{q}$ has been expanded and truncated at first order in $\chi$. Moreover, we also have
\begin{equation}
{\bar{t}}(r) =-{\sqrt{q}} \int \frac{\mathrm{d}r} {f(r)\sqrt{1-f(r)}}
\end{equation}
and considering up to the first order in $\chi$, we obtain
\begin{equation}
{\bar{t}}(r) \approx \sqrt{q}\left[ 4M\tanh^{-1} \left( \sqrt{2}\sqrt{\frac{M}{r}} \right) -\frac{1}{3} \sqrt{2}\sqrt{\frac{r}{M}}(6M+r) \right] +C'.
\end{equation}

Here, $C$ and $C'$ denote arbitrary integration constants. The emitted radiation is modeled as a massless scalar field with wave function $\Psi$, whose dynamics in the curved background obey the covariant Klein--Gordon equation,
\begin{equation}
\frac{1}{\sqrt{-g}} \partial_\mu \left( \sqrt{-g}\,g^{\mu\nu}\partial_\nu \right) \Psi_{\omega_{\infty}} ({\bar{t}},r) =0.
\end{equation}
The scalar mode of frequency $\omega_{\infty}$ admits a separable representation $\Psi_{\omega_{\infty}}({\bar{t}},r)$ adapted to the given spacetime
\begin{equation}
\Psi_{\omega_{\infty}} ({\bar{t}},r) \sim e^{ -i\omega_{\infty}{\bar{t}} +i\omega_{\infty}{\bar{r}}_*(r) }.
\end{equation}
The quantity $\omega_{\infty}$ labels the Killing frequency conjugate to the {normalized} coordinate ${\bar{t}}$. The frequency measured with respect to the proper time of a static observer at spatial infinity is {therefore the same quantity}. To isolate the atomic excitation process from background Hawking emission, the analysis adopted a configuration in which a reflecting boundary encloses the black hole. This construction suppresses outgoing thermal radiation and effectively realizes a Boulware-type vacuum state. Within this framework, the internal atomic dynamics were described through the lowering operator $\varsigma=|g\rangle\langle e|$, which mediates transitions between the excited and ground levels. The interaction between the atom and the scalar field was then encoded in the corresponding atom field Hamiltonian
\begin{equation}
\label{atom_field_hamiltonian}\hat{\mathrm{H}}(\tau) = \hbar\mathfrak{a} \left[ \hat{b} \Psi_{\omega_{\infty}} ({\bar{t}}(\tau),r(\tau)) +\text{h.c.} \right] \left[ \varsigma e^{-i\tilde{\omega}\tau} +\text{h.c.} \right].
\end{equation}

In Eq.~({\hypersetup{citecolor=red,linkcolor=red}\ref{atom_field_hamiltonian}}), the parameter $\mathfrak{a}$ characterizes the strength of the coupling between the atomic degrees of freedom and the scalar field, while $\hat{b}$ denotes the annihilation operator. The quantity $\tilde{\omega}$ represents the intrinsic transition frequency of the two--level atom.


{\subsection{{HBAR master equation and entropy flux}}\label{sec:hbar_formalism}}

{{Following the} HBAR {\hypersetup{citecolor=red,linkcolor=red}construction of Scully et al.~\cite{scully2018quantum}, we use the reduced field master equation and its standard stationary distribution. This} construction describes the entropy carried by acceleration radiation generated through the interaction of a stream of freely falling two--level atoms with the quantum field near the event horizon \cite{scully2018quantum}. Let $\varrho_{\omega_\infty}$ denote the reduced density matrix of a field mode with asymptotic frequency $\omega_\infty$, after the atomic degrees of freedom have been traced out. We assume that the atoms cross the interaction region independently, with an injection rate $\mathcal{R}=\mathrm{d}N/\mathrm{d}\bar t$ measured with respect to the normalized asymptotic time $\bar t$. In the weak--coupling and Markovian regimes, the change produced during an interval $\Delta\bar t$ by $\Delta N=\mathcal{R}\Delta\bar t$ atoms is}
\begin{equation}
{
\begin{aligned}
\varrho_{\omega_\infty}(\bar t+\Delta\bar t) -\varrho_{\omega_\infty}(\bar t) =\Delta N\Big[ &\mathrm{P}_{\mathrm{abs}}\, \mathcal{D}[\hat b]\varrho_{\omega_\infty} +\mathrm{P}_{\mathrm{exc}}\, \mathcal{D}[\hat b^\dagger]\varrho_{\omega_\infty} \Big],
\end{aligned}}
\label{singleatommap}
\end{equation}
{where $\hat b$ and $\hat b^\dagger$ are the annihilation and creation operators of the selected field mode and $P_{\mathrm{exc}}\equiv P_{g,0\rightarrow e,1}^{\mathrm{emi}},\qquad P_{\mathrm{abs}}\equiv P_{g,1\rightarrow e,0}${. The transition probabilities entering these coefficients are evaluated} in {\hypersetup{citecolor=red,linkcolor=red}Sec.~\ref{sec:atomic_results}. In} addition, }
\begin{equation}
{ \mathcal{D}[\hat L]\varrho = \hat L\varrho\hat L^\dagger -\frac{1}{2} \left\{ \hat L^\dagger\hat L,\varrho \right\} }
\label{lindbladdissipator}
\end{equation}
{is the Lindblad dissipator. Dividing Eq.~(\ref{singleatommap}) by $\Delta\bar t$ and taking the continuous limit gives the reduced master equation, in the interaction picture,}
\begin{equation}
{ \dot{\varrho}_{\omega_\infty} = \Gamma_{\mathrm{abs}} \mathcal{D}[\hat b]\varrho_{\omega_\infty} + \Gamma_{\mathrm{exc}} \mathcal{D}[\hat b^\dagger]\varrho_{\omega_\infty}, \qquad \Gamma_{\mathrm{abs}} =\mathcal{R}\mathrm{P}_{\mathrm{abs}}, \qquad \Gamma_{\mathrm{exc}} =\mathcal{R}\mathrm{P}_{\mathrm{exc}}. }
\label{hbarlindblad}
\end{equation}
{Here and below, an overdot denotes a derivative with respect to $\bar t$. Projecting Eq.~(\ref{hbarlindblad}) onto the number basis, $\varrho_{n,n}=\langle n|\varrho_{\omega_\infty}|n\rangle$, yields}
\begin{equation}
{
\begin{aligned}
\dot{\varrho}_{n,n} = &\Gamma_{\mathrm{abs}} \left[ (n+1)\varrho_{n+1,n+1} -n\varrho_{n,n} \right] \\ &+ \Gamma_{\mathrm{exc}} \left[ n\varrho_{n-1,n-1} -(n+1)\varrho_{n,n} \right].
\end{aligned}}
\label{populationmaster}
\end{equation}
{The stationary solution is obtained by imposing $\dot{\varrho}_{n,n}=0$ in Eq.~(\ref{populationmaster}). For $\Gamma_{\mathrm{exc}}<\Gamma_{\mathrm{abs}}$, normalization gives the geometric distribution}
\begin{equation}
{ \varrho_{n,n}^{\mathrm{st}} = \left( 1-\frac{\Gamma_{\mathrm{exc}}}{\Gamma_{\mathrm{abs}}} \right) \left( \frac{\Gamma_{\mathrm{exc}}}{\Gamma_{\mathrm{abs}}} \right)^n, \qquad \sum_{n=0}^{\infty}\varrho_{n,n}^{\mathrm{st}}=1. }
\label{vvcccccxx}
\end{equation}
{The superscript $\mathrm{st}$ denotes the instantaneous stationary reference distribution. The entropy production calculated below describes a slow departure from this reference state and therefore does not require $\dot{\varrho}_{n,n}$ to vanish identically.}

The entropy of the system is quantified through the von Neumann definition, constructed from the reduced density matrix that characterizes the {field state after the atomic degrees of freedom have been traced out}. In other words, it reads
\begin{equation}
S_\varrho = -k_{B} \sum\limits_{n,\omega_{\infty}} \varrho_{n,n} \ln(\varrho_{n,n}).
\end{equation}
Moreover, the entropy production rate associated with the emission of real scalar modes is then expressed as
\begin{equation}
\dot{S}_\varrho = -k_{B} \sum\limits_{n,\omega_{\infty}} \dot{\varrho}_{n,n} \ln(\varrho_{n,n})\,.
\end{equation}
{The term proportional to $\sum_n\dot{\varrho}_{n,n}$ vanishes because the evolution preserves $\mathrm{Tr}(\varrho)=1$.} {In the quasi-stationary regime, using} the {stationary reference} solution of the density matrix, the {standard quasi stationary} entropy {flux} can be expressed as
\begin{equation}\label{2.46}
\dot{S}_\varrho \approx -k_{B} \sum\limits_{n,\omega_{\infty}} \dot{\varrho}_{n,n} \ln(\varrho^{{\mathrm{st}}}_{n,n})\,.
\end{equation}


{\subsection{{Canonical energy and black hole entropy}}\label{sec:canonical}}

{\hypersetup{citecolor=red,linkcolor=red}The canonical energy and thermodynamic entropy used here are established properties of the adopted black hole solution~\cite{Liu:2025oho}. In the notation of that reference}, {$q=1+\ell_1+\ell_2$, with $\ell_2=\beta^2\ell$ and $\ell_1=\ell$} for the {spacelike vacuum or $\ell_1=0$ for the lightlike vacuum. Thus $q_{\mathrm S}=1+(\beta^2+1)\ell$ and $q_{\mathrm L}=1+\beta^2\ell$}. The {superscripts $\mathrm S$ and $\mathrm L$ identify these two vacuum configurations.

The geometry} is asymptotically conical. {For} the normalized Killing field {$\partial_{\bar t}$}, the canonical conserved energy {\hypersetup{citecolor=red,linkcolor=red}$\mathcal M_\infty\equiv E_W$ is obtained from Ref.~\cite{Liu:2025oho}}, {specialized} to the {neutral solution:}
\begin{equation}
{
\begin{aligned}
\mathcal{M}_\infty^{(\mathrm{S})} \equiv E_W^{(\mathrm{S})} &= \frac{1+\ell}{\sqrt{q_{\mathrm{S}}}}\,M, \qquad q_{\mathrm{S}}=1+(\beta^2+1)\ell, \\ \mathcal{M}_\infty^{(\mathrm{L})} \equiv E_W^{(\mathrm{L})} &= \frac{1}{\sqrt{q_{\mathrm{L}}}}\,M, \qquad q_{\mathrm{L}}=1+\beta^2\ell.
\end{aligned}}
\label{canonicalmass}
\end{equation}

{Equivalently}, {$\mathcal M_\infty=(1+\ell_1)M/\sqrt q$}. {The horizon area is $A_H=4\pi r_h^2=16\pi M^2$}. {At} fixed $\ell$ and $\beta$, the first law {with this energy and the normalized temperature} gives {\hypersetup{citecolor=red,linkcolor=red}the known thermodynamic entropies~\cite{Liu:2025oho}}
\begin{equation}
{ S_{\mathrm{th}}^{(\mathrm{S})} = (1+\ell)\frac{k_B A_H}{4}, \qquad S_{\mathrm{th}}^{(\mathrm{L})} = \frac{k_B A_H}{4}. }
\label{branchthermodynamicentropy}
\end{equation}

{\hypersetup{citecolor=red,linkcolor=red}The corresponding Wald entropies are $S_W^{(\mathrm S)}=(1+\ell/2)k_BA_H/4$ and $S_W^{(\mathrm L)}=k_BA_H/4$~\cite{Liu:2025oho}. These energy and entropy formulas are background inputs. Section~\ref{sec:hbar_results} uses them to express} the HBAR {entropy flux in terms of radiative} area {loss}. {We keep $k_B$ explicit} in the {entropy formulas; temperatures written in frequency units satisfy $k_BT_H=\bar\kappa/(2\pi)$}.


{\section{{Results}}\label{sec:results}}

{\hypersetup{citecolor=red,linkcolor=red}We now apply the ingredients of Sec.~\ref{sec:background} to the bumblebee geometry. The results concern the geometric dependence of the scalar correlations, the normalized excitation probability of the falling atom, and the HBAR entropy flux and area coefficient. The standard correlation measures, thermal factor, master equation, and canonical black hole quantities enter as the explicitly cited inputs.}


{\subsection{{Entanglement and mutual information}}\label{sec:correlation_results}}

\begin{figure}[h]
\centering
\includegraphics[width=0.32\linewidth]{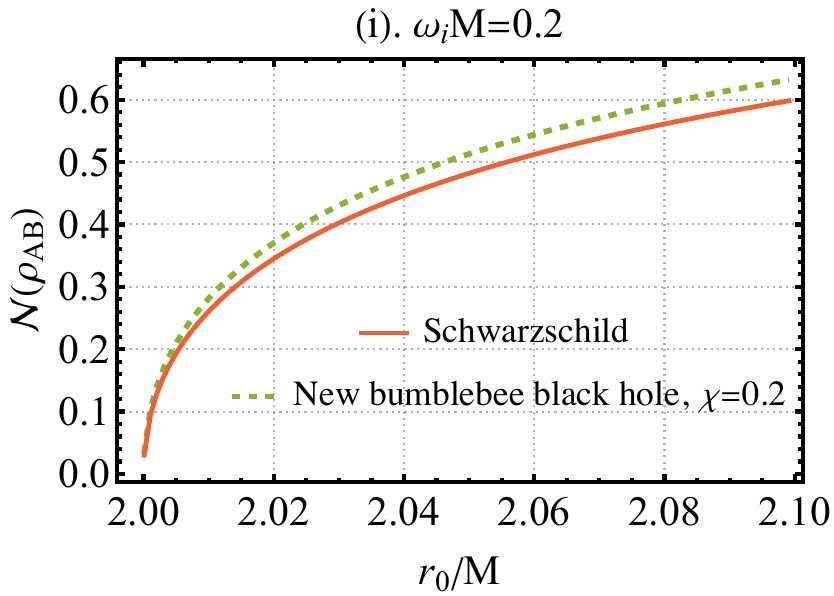}
\includegraphics[width=0.32\linewidth]{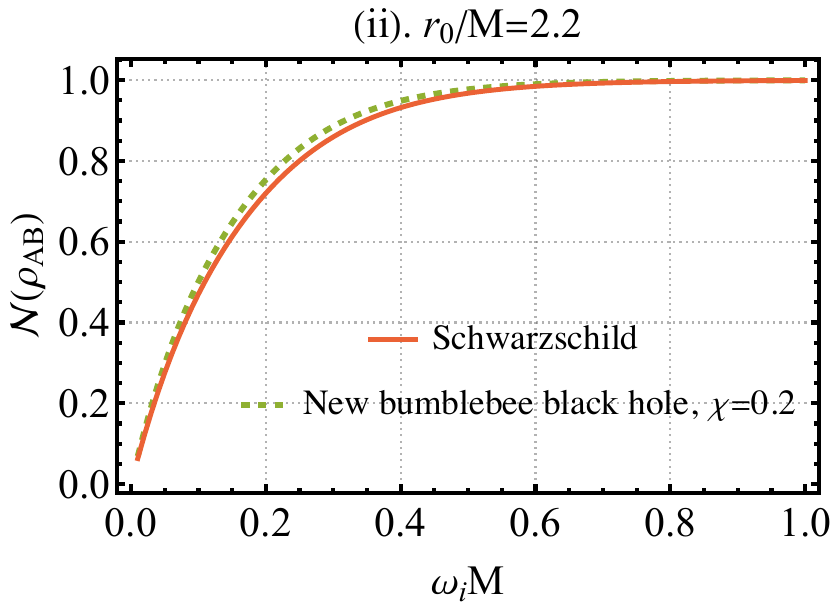}
\includegraphics[width=0.32\linewidth]{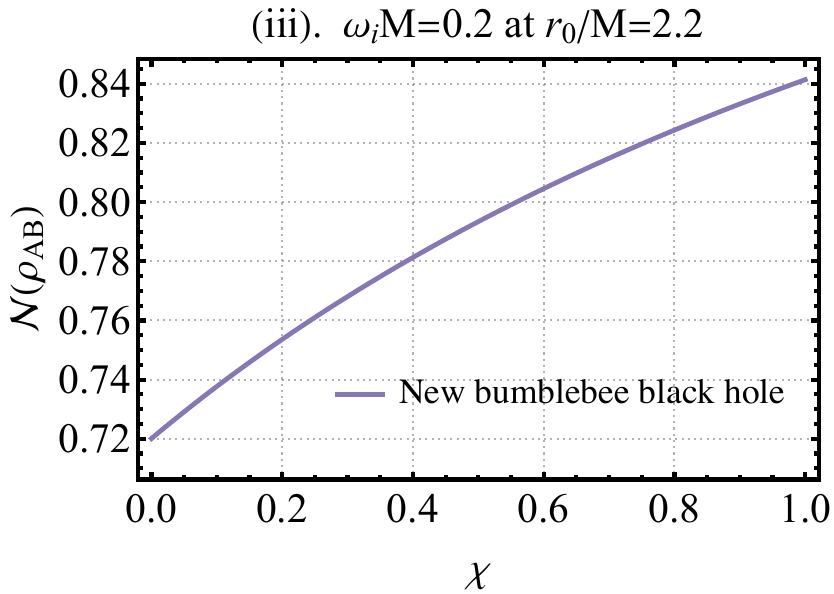}
\includegraphics[width=0.32\linewidth]{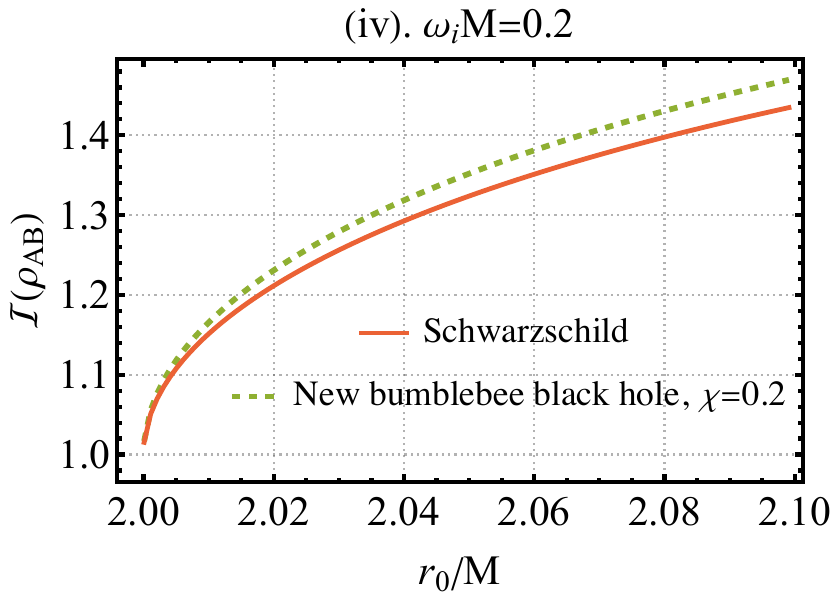}
\includegraphics[width=0.32\linewidth]{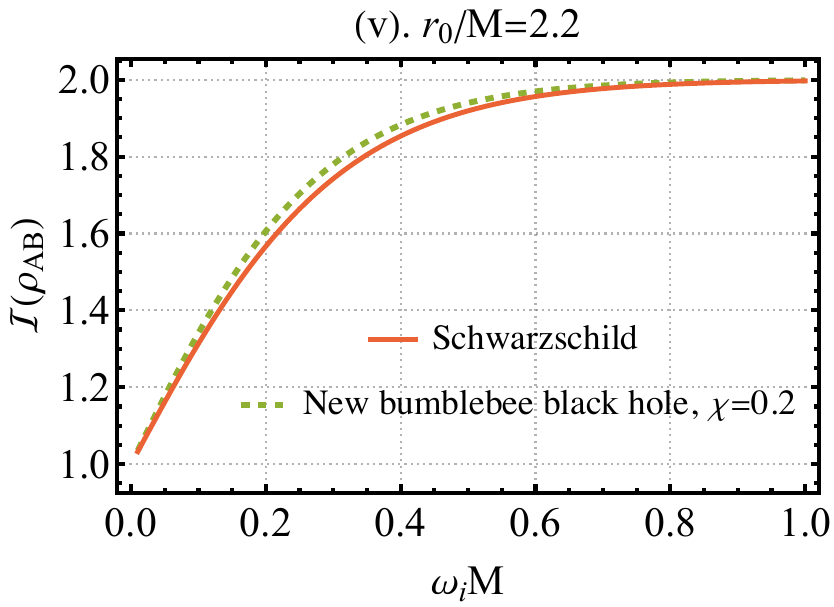}
\includegraphics[width=0.32\linewidth]{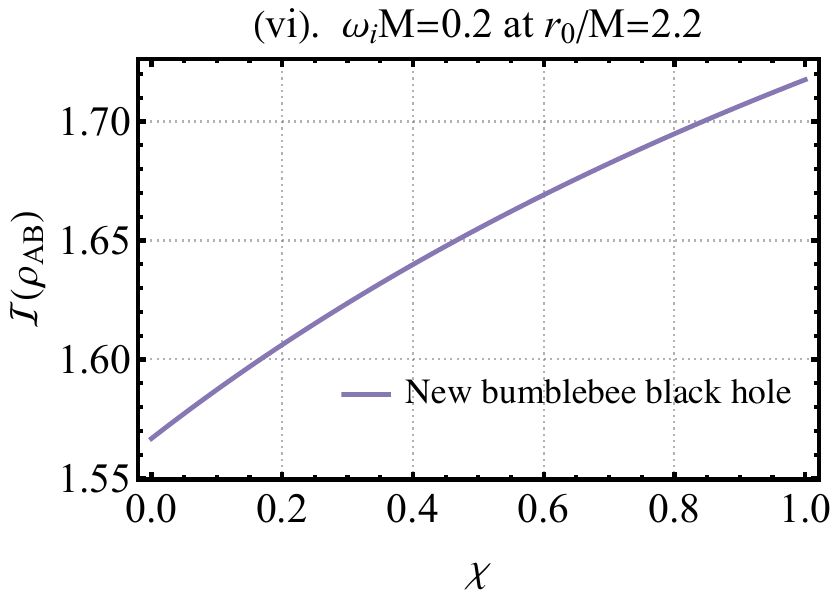}
\caption{{{Logarithmic} negativity {$\mathcal N(\rho_{\mathrm{AB}})$ in panels (i), (ii), and (iii),} and mutual information {$I(\rho_{\mathrm{AB}})$ in panels (iv), (v), and (vi). Panels (i) and (iv) vary} Bob's position {$r_0/M$ at fixed $\omega_iM=0.2$; panels (ii) and (v) vary} the {local} packet frequency {$\omega_iM$ at fixed $r_0/M=2.2$. These four panels compare Schwarzschild spacetime ($\chi=0$) with the bumblebee geometry at $\chi=0.2$. Panels (iii) and (vi) vary $\chi=\alpha\ell$ at fixed $\omega_iM=0.2$ and $r_0/M=2.2$}.}}
\label{fig2}
\end{figure}

\begin{figure}[h]
\centering
\includegraphics[width=0.32\linewidth]{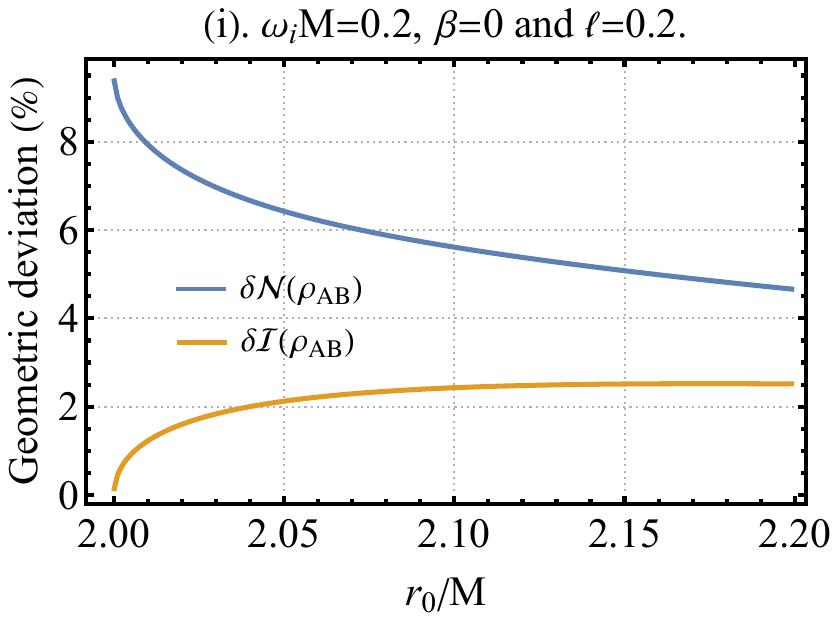}
\includegraphics[width=0.32\linewidth]{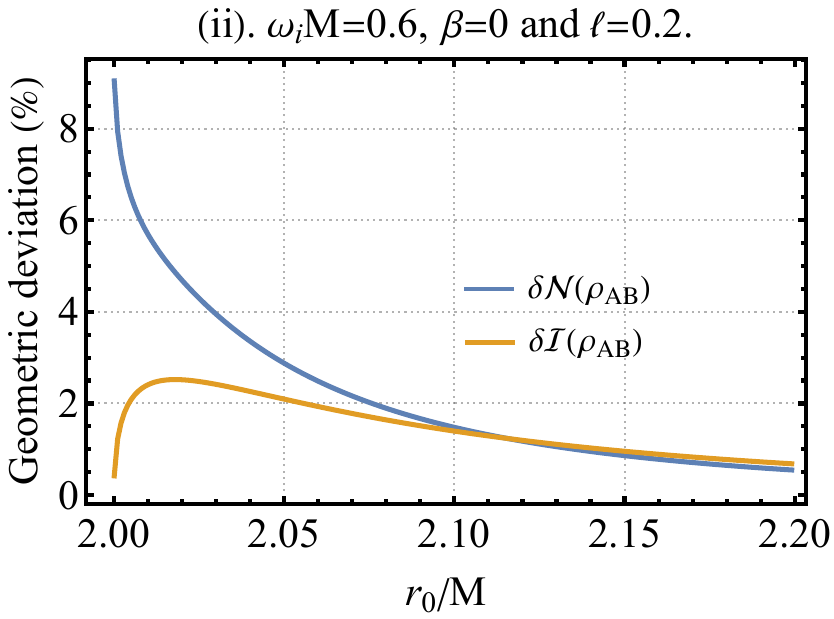}
\includegraphics[width=0.32\linewidth]{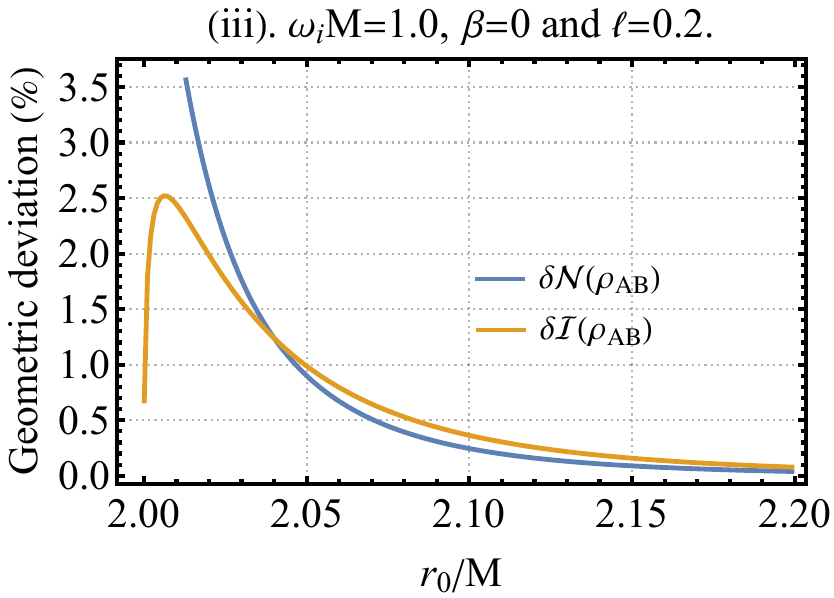}
\includegraphics[width=0.32\linewidth]{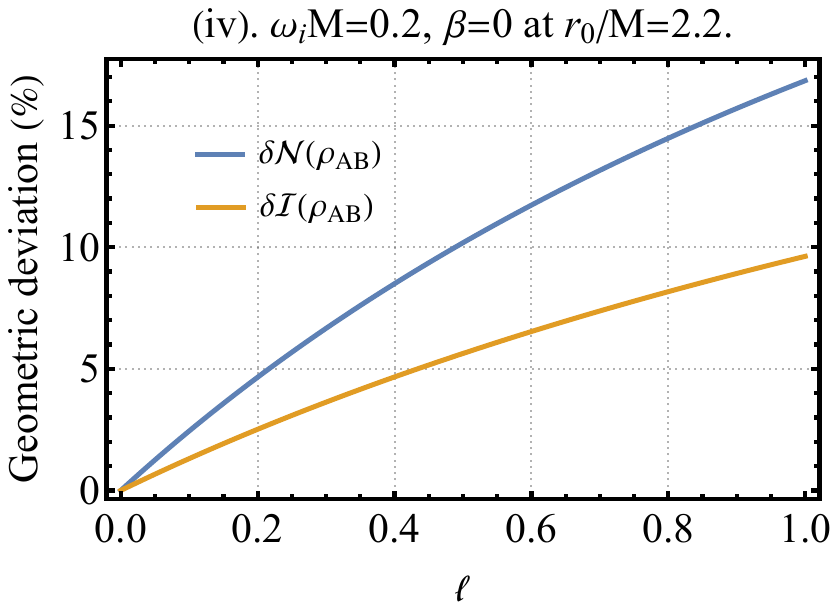}
\includegraphics[width=0.32\linewidth]{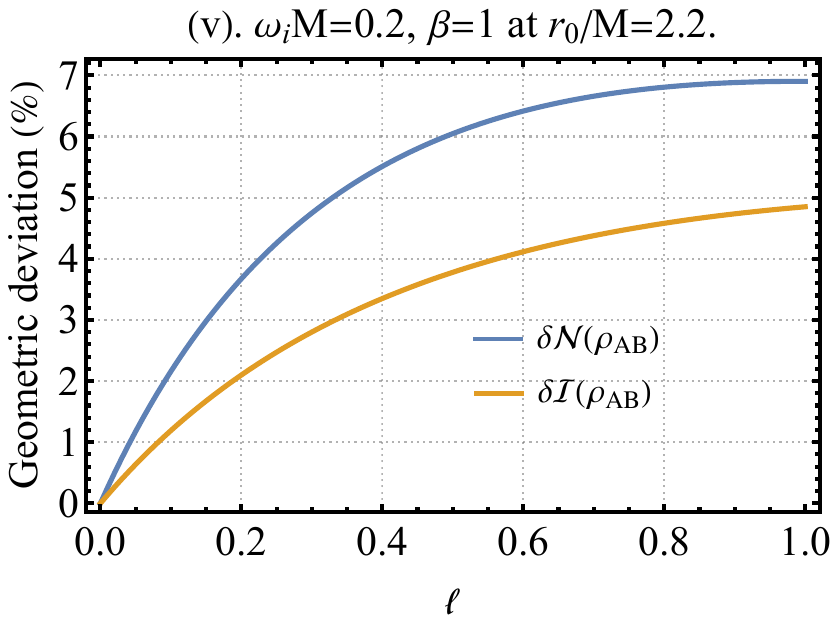}
\includegraphics[width=0.32\linewidth]{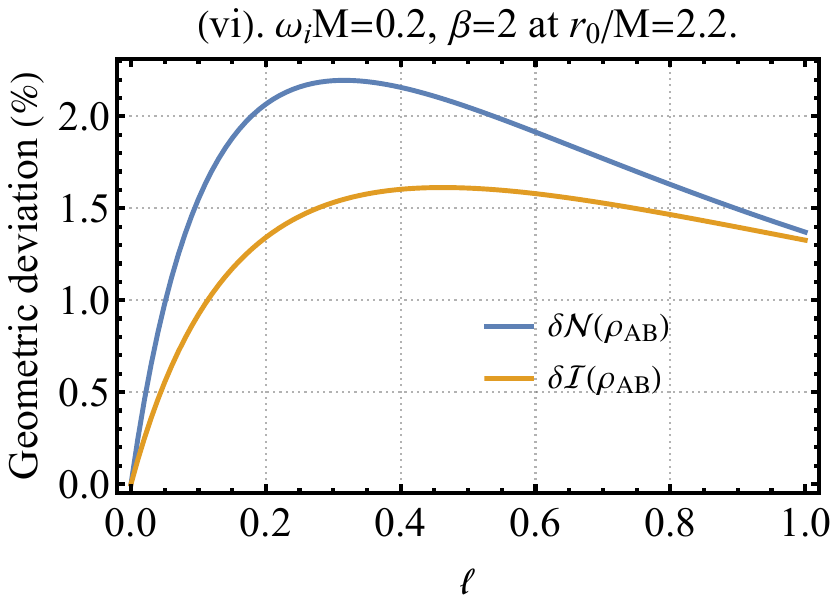}
\includegraphics[width=0.32\linewidth]{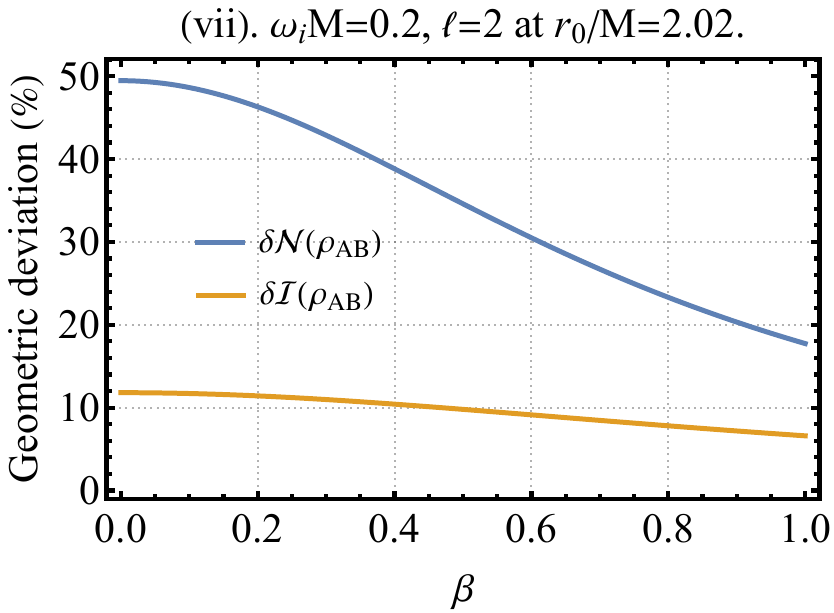}
\includegraphics[width=0.32\linewidth]{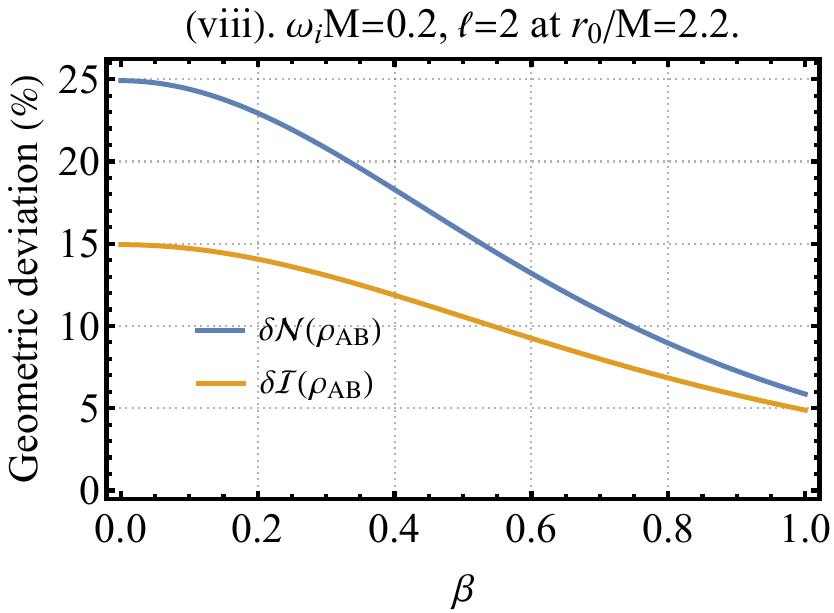}
\includegraphics[width=0.32\linewidth]{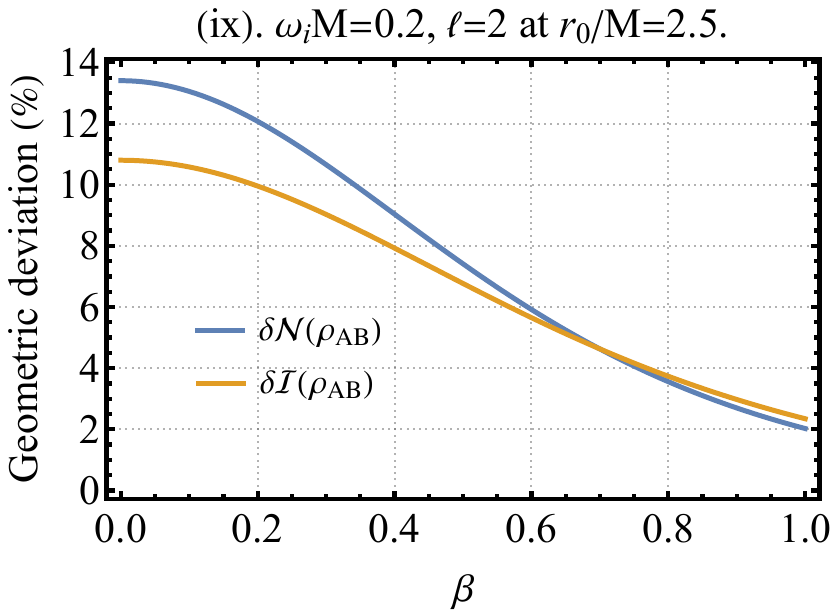}
\caption{{Relative {deviations $\delta\mathcal N$ (blue)} and {$\delta I$ (orange)} between spacelike and lightlike {vacuum configurations. Panels (i)}, {(ii)}, and {(iii) vary $r_0/M$} at fixed {$\ell=0.2$ and $\beta=0$, with $\omega_iM=0.2$, $0.6$, and $1.0$, respectively. Panels (iv), (v), and (vi) vary $\ell$ at fixed $\omega_iM=0.2$ and $r_0/M=2.2$, with $\beta=0$, $1$, and $2$, respectively. Panels (vii), (viii), and (ix) vary $\beta$ at fixed $\ell=2$ and $\omega_iM=0.2$, with $r_0/M=2.02$, $2.2$, and $2.5$, respectively. Each comparison uses the same} $\beta$ and $\ell$ {for both vacuum configurations}, {so that $\chi_{\mathrm S}-\chi_{\mathrm L}=\ell$}.}}
\label{fig3}
\end{figure}

{Positive geometric deformation partially counteracts the loss of scalar field correlations, and spacelike and lightlike vacuum configurations with the same metric give identical logarithmic negativity and mutual information.} We {obtain this result by evaluating the established measures in} Eqs.~(\ref{NS1}) and (\ref{IS1}) {with} the {\hypersetup{citecolor=red,linkcolor=red}geometry dependent squeezing parameter in Eq.~(\ref{tansi}). The resulting parameter} dependence {\hypersetup{citecolor=red,linkcolor=red}and the comparisons in Figs.~\ref{fig2} and \ref{fig3} are the correlation results} of the {present work}.

{At fixed $\beta$ and $\ell$, the spacelike and lightlike parametrizations give}
\begin{equation}\label{branch_chi_ent}
{ \chi_{\rm S}=(\beta^2+1)\ell, \qquad \chi_{\rm L}=\beta^2\ell, \qquad \chi_{\rm S}-\chi_{\rm L}=\ell. }
\end{equation}
To compare the two {configurations}, we introduce a deviation function, exemplified by the logarithmic negativity,
\begin{equation}
\delta\mathcal{N}
=\frac{\mathcal{N}_{\rm S}-\mathcal{N}_{\rm L}}{\mathcal{N}_{\rm L}}\times 100\%,
\end{equation}
which quantifies the relative difference in entanglement between the spacelike and lightlike Lorentz--violating {configurations}. {The same definition is used for $\delta I$. Equation~(\ref{branch_chi_ent}) shows that this comparison involves two different values of $\chi$ and, therefore, two different metric deformations. If $\chi$ is held fixed, the scalar correlation measures are identical and both deviations vanish.}

Before comparing the two {configurations}, we first briefly examine the impact of Lorentz violation itself on entanglement degradation. Panels (i)–(iii) in Fig. \ref{fig2} summarize the behavior of the logarithmic negativity $\mathcal{N}(\rho_{\rm AB})$ in the new bumblebee black hole spacetime. Panel (i) shows $\mathcal{N}(\rho_{\rm AB})$ as a function of the observer's radial position $r_0/M$ for a fixed frequency $\omega_i M = 0.2$. As Bob approaches the event horizon ($r_0 \to 2M$), the logarithmic negativity decreases rapidly{, as expected from the increasing thermal mixing between the exterior packet and its interior partner}. {For $\alpha\ell=0.2$, the dashed curve remains above the Schwarzschild curve. At fixed $r_0$ and $\omega_i$, a positive value of $\alpha\ell$ lowers the local Hawking temperature and leaves a larger residual entanglement.}

Panel (ii) displays the dependence of $\mathcal{N}(\rho_{\rm AB})$ on the {central packet frequency} $\omega_i M$ for a fixed radial position $r_0/M = 2.2$. The logarithmic negativity increases monotonically with $\omega_i$, indicating that low--frequency {packets} are strongly affected by horizon-induced noise, whereas higher-frequency {packets} retain a larger fraction of their initial entanglement. {The curve for $\alpha\ell=0.2$ remains above the Schwarzschild curve, and their difference is largest at low frequency. At high frequency both curves approach the value of the initial Bell state.} Finally, panel (iii) isolates the effect of the Lorentz--violating parameter by plotting $\mathcal{N}(\rho_{\rm AB})$ as a function of the effective combination $\alpha\ell$ at fixed $\omega_i M = 0.2$ and $r_0/M = 2.2$. The logarithmic negativity increases monotonically with $\alpha\ell$, leading to an overall enhancement of {approximately $17\%$} as $\alpha\ell$ is raised from zero (the Schwarzschild limit) to unity. In addition, panels (iv)–(vi) in Fig.~\ref{fig2} display the behavior of the mutual information under the same conditions. {The mutual information has the same general dependence on $r_0$, $\omega_i$, and $\alpha\ell$.}

{We discuss the deviation functions in Fig.~\ref{fig3} row by row.}
Panels (i)–(iii) display the relative deviations $\delta\mathcal{N}(\rho_{\rm AB})$ and $\delta {I}(\rho_{\rm AB})$ as functions of the radial position $r_0/M$ for increasing {packet frequencies} $\omega_i M = 0.2,\,0.6,$ and $1.0$, with $\ell=0.2$ and $\beta=0$ fixed.
{For the three frequencies, $\delta\mathcal{N}$ is largest close to the horizon and decreases as Bob moves outward. The relative mutual information has a weaker response and is nonmonotonic for the two larger frequencies. These results show that the two measures do not have the same dependence on $r_0$.} Panels (iv)–(vi) illustrate the effect of the Lorentz--violating parameter $\ell$ at fixed $\omega_i M = 0.2$ and $r_0/M = 2.2$ for three representative values of $\beta$.
{For $\beta=0$, both deviations increase with $\ell$. For $\beta=1$, their growth becomes progressively weaker, whereas for $\beta=2$ a maximum occurs at an intermediate value of $\ell$.}

Panels (vii)–(ix) show the variation of the deviations with respect to $\beta$ for fixed $\ell=2$ and three values of the Bob position ${r_0/M}$. In all cases, $\delta\mathcal{N}(\rho_{AB})$ and $\delta\mathcal{I}(\rho_{AB})$ decrease monotonically with increasing $\beta${. This occurs because the common contribution $\beta^2\ell$ becomes large compared with the fixed difference $\chi_{\rm S}-\chi_{\rm L}=\ell$}.



{\subsection{{Response of the infalling atom}}\label{sec:atomic_results}}

{\hypersetup{citecolor=red,linkcolor=red}At leading near horizon order, the freely falling atom has a Planckian excitation probability governed by the normalized surface gravity; its local thermal factor agrees with the flat spacetime acceleration radiation expression under $\omega_\infty,\bar\kappa\ll\tilde\omega$. We derive the bumblebee parameter dependence using the trajectory and interaction Hamiltonian of Sec.~\ref{sec:atomic_formalism} and the established acceleration radiation method~\cite{scully2018quantum,sen2022equivalence}.}

The corresponding excitation probability follows directly from the transition amplitude associated with this interaction Hamiltonian and is expressed as
\begin{equation}
\mathrm{P}_{g,0\rightarrow e,1} = \frac{1}{\hbar^2} \left| \int\mathrm{d}\tau \langle1_{\omega_{\infty}},e| \hat{\mathrm{H}}(\tau) |0_{\omega_{\infty}},g\rangle \right|^2.
\end{equation}
Since $\omega_{\infty}$ is already the Killing frequency appearing in the {normalized} field mode, no {additional redshift or} position dependent ``frequency at infinity'' is introduced in the transition amplitude. Selecting the excitation with emission term and changing the integration variable from $\tau$ to $r$ gives
\begin{equation}\label{mnnnn}
\begin{split}
\mathrm{P}_{g,0\rightarrow e,1} &= \mathfrak{a}^2 \left| \int_\infty^{r_h} \mathrm{d}r\, \frac{\mathrm{d}\tau}{\mathrm{d}r} e^{ i\omega_{\infty} {\bar{t}}(r) -i\omega_{\infty} {\bar{r}}_*(r) } e^{i\tilde{\omega}\tau(r)} \right|^2 \\ &= \mathfrak{a}^2 \left| \int_{r_h}^{\infty} \mathrm{d}r\, \frac{\sqrt{q}} {\sqrt{1-f(r)}} e^{ -i\omega_{\infty} {\sqrt{q}} \int \frac{\mathrm{d}r} {f(r)\sqrt{1-f(r)}} -i\omega_{\infty} {\sqrt{q}} \int \frac{\mathrm{d}r} {f(r)} } e^{ -i\tilde{\omega}\sqrt{q} \int \frac{\mathrm{d}r} {\sqrt{1-f(r)}} } \right|^2.
\end{split}
\end{equation}
The explicit dependence of the excitation probability follows by expressing the proper time $\tau$, the coordinate time ${\bar{t}}$, and the tortoise coordinate ${\bar{r}}_*$ as functions of the radial variable. Using the near-horizon expansion {$f(r)=f'(r_h)x+\mathcal{O}(x^2)$} and defining $x=r-r_h$, Eq.~(\ref{mnnnn}) reduces at leading order to
\begin{equation}
\mathrm{P}_{g,0\rightarrow e,1} \simeq q\mathfrak{a}^2 \left| \int_0^\infty \mathrm{d}x\, x^{-{\frac{i\omega_\infty}{\bar{\kappa}}}} e^{-i\sqrt{q}\,\tilde{\omega}x} \right|^2.
\label{nearhorizonamplitudeprecise}
\end{equation}
The upper limit has been extended in the standard near horizon approximation, which is justified when ${\omega_\infty,\bar{\kappa}}\ll\tilde{\omega}$. With $y=\sqrt{q}\,\tilde{\omega}x$, the factors of $q$ cancel from the prefactor, and
\begin{equation}
\mathrm{P}_{g,0\rightarrow e,1} \simeq \frac{\mathfrak{a}^2}{\tilde{\omega}^2} \left| \int_0^\infty \mathrm{d}y\, y^{-{\frac{i\omega_\infty}{\bar{\kappa}}}} e^{-iy} \right|^2.
\label{dimensionlessamplitudeprecise}
\end{equation}
Since both ${\omega_\infty}$ and ${\bar{\kappa}}$ remain much smaller than $\tilde{\omega}$ in natural units, the excitation probability simplifies substantially. Under this hierarchy of scales, the standard Gamma function identity gives
\begin{equation}
\mathrm{P}^{\mathrm{emi}}_{g,0\rightarrow e,1} \cong \frac{ {2}\pi \mathfrak{a}^2 \omega_{\infty} } { {\bar{\kappa}} \tilde{\omega}^2 } \frac{1} { e^{ \frac{ {2}\pi \omega_{\infty} }{ {\bar{\kappa}} } } -1 }.
\label{emissionprecise}
\end{equation}
The corresponding absorption probability is
\begin{equation}
\mathrm{P}^{\mathrm{abs}}_{e,0\rightarrow g,1} \cong \frac{ {2}\pi \mathfrak{a}^2 \omega_{\infty} } { {\bar{\kappa}} \tilde{\omega}^2 } \frac{1} { 1- e^{ -\frac{ {2}\pi \omega_{\infty} }{ {\bar{\kappa}} } } }.
\label{absorptionprecise}
\end{equation}
At this stage, {using $\bar{\kappa}=1/(4M\sqrt{q})$}, the excitation probability can be expressed directly in terms of the frequency measured by an observer at infinity, $\omega_\infty$, leading to the following representation
\begin{equation}
\mathrm{P}^{\mathrm{emi}}_{g,0\rightarrow e,1} \cong \frac{ 8\pi M\sqrt{q}\, \mathfrak{a}^2\omega_\infty } {\tilde{\omega}^2} \frac{1} { e^{8\pi M\sqrt{q}\,\omega_\infty}-1 }.
\label{infinityfrequencyprecise}
\end{equation}
For the local comparison, the {normalized} frequency and the Rindler acceleration introduced in Sec.~{\hypersetup{citecolor=red,linkcolor=red}\ref{sec:rindler}} satisfy
$\omega_{\infty}/{\bar{\kappa}}
=\omega_{\mathrm{loc}}/a_R$,
{where $\omega_{\mathrm{loc}}=\omega_\infty/\sqrt{f_0}$
and $a_R=\bar{\kappa}/\sqrt{f_0}$}.
Eq.~(\ref{emissionprecise}) becomes
\begin{equation}
\mathrm{P}^{\mathrm{emi}}_{g,0\rightarrow e,1} \cong \frac{ 2\pi\mathfrak{a}^2 \omega_{\mathrm{loc}} } { a_R\tilde{\omega}^2 } \frac{1} { e^{ \frac{2\pi\omega_{\mathrm{loc}}}{a_R} } -1 }.
\label{localresponseprecise}
\end{equation}
{The Planck factor is} the {\hypersetup{citecolor=red,linkcolor=red}familiar acceleration radiation factor~\cite{scully2018quantum,sen2022equivalence}. Here} the {result is its explicit dependence on the bumblebee geometry through $\bar\kappa$ and the equality} of the {local thermal factors after matching $\omega_{\mathrm{loc}}/a_R$. This comparison applies} at leading near horizon order {under} the {stated frequency hierarchy and supports} the {local} equivalence principle {interpretation within those approximations}.


{\subsection{{HBAR spectrum, entropy flux, and area relation}}\label{sec:hbar_results}}

\begin{figure}
\centering
\includegraphics[scale=0.51]{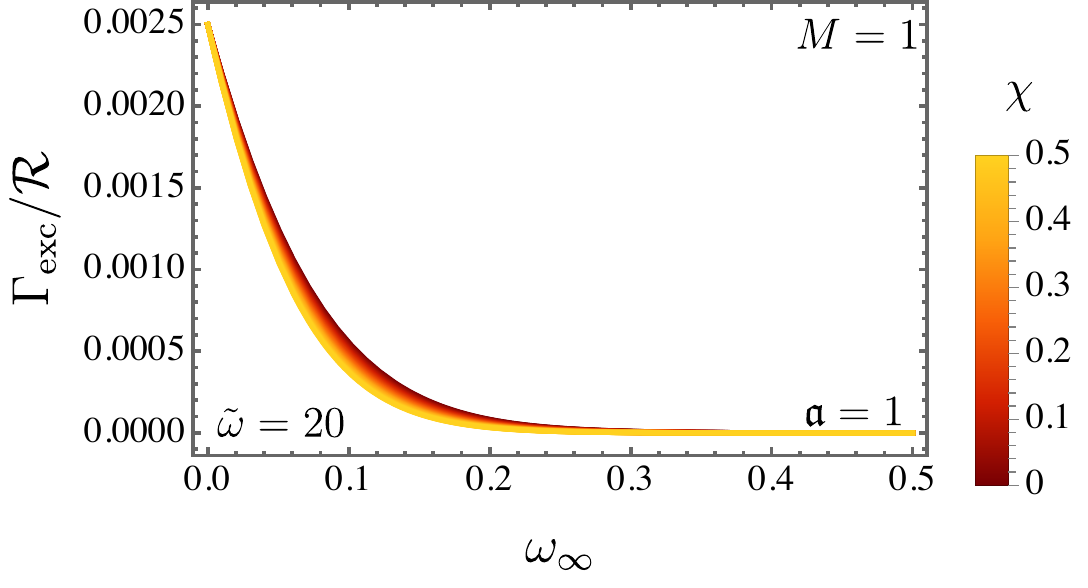}
\caption{${\Gamma_{\mathrm{exc}}/\mathcal{R}=\mathrm{P}_{\mathrm{exc}}}$ is shown as a function of $\omega_{\infty}$ for different values of $\chi${, with $M=1$, $\mathfrak{a}=1$, and $\tilde{\omega}=20$}.}
\label{gammaexs}
\end{figure}

\begin{figure}
\centering
\includegraphics[scale=0.51]{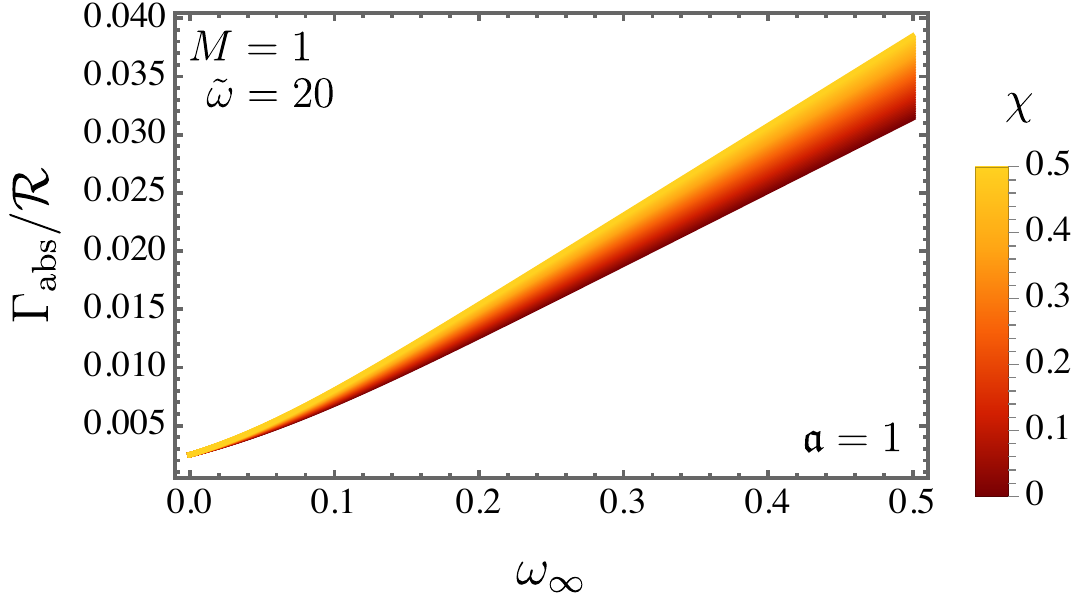}
\caption{${\Gamma_{\mathrm{abs}}/\mathcal{R}=\mathrm{P}_{\mathrm{abs}}}$ is shown as a function of $\omega_{\infty}$ for different values of $\chi${, with $M=1$, $\mathfrak{a}=1$, and $\tilde{\omega}=20$}.}
\label{gammaabs}
\end{figure}

The {\hypersetup{citecolor=red,linkcolor=red}HBAR entropy flux is proportional to $\sqrt q\,\dot E_{\mathrm{rad}}$, while its area coefficient becomes $(k_B/4)(1+\ell)$ for the spacelike vacuum and $k_B/4$ for the lightlike vacuum after the canonical energy is used. We derive these geometry dependent results from the standard master equation and entropy flux of Sec.~\ref{sec:hbar_formalism} together with the established thermodynamic inputs in Sec.~\ref{sec:canonical}.

\subsubsection{{Spectral rates and detailed balance}}

Positive $\chi$ suppresses the normalized excitation rate and enhances the normalized absorption rate, while their} ratio {\hypersetup{citecolor=red,linkcolor=red}follows the thermal detailed balance factor set by $\bar\kappa$.

The established detailed balance relation~\cite{scully2018quantum},} evaluated {\hypersetup{citecolor=red,linkcolor=red}with the probabilities in Sec.~\ref{sec:atomic_results}, gives the explicit bumblebee dependence} under the assumption {$\omega_\infty,\bar{\kappa}\ll\tilde{\omega}$}, which simplifies its analytical form
\begin{equation}
\frac{\Gamma_{\mathrm{exc}}} {\Gamma_{\mathrm{abs}}} = { \frac{\mathrm{P}_{\mathrm{exc}}} {\mathrm{P}_{\mathrm{abs}}}} = e^{-8\pi M{\sqrt{q}\,\omega_\infty}}.
\label{ratebalance}
\end{equation}
{Here, $\mathrm{P}_{\mathrm{abs}}$ denotes the coefficient of the field loss channel in the master equation. The probability evaluated through the time reversed atomic transition in {\hypersetup{citecolor=red,linkcolor=red}Sec.~\ref{sec:atomic_results}} gives the same coefficient by microreversibility. The detailed-balance ratio in Eq.~(\ref{ratebalance}) is independent of the common atomic infall rate $\mathcal{R}$ and of the overall coupling normalization.} To clarify the physical content of the results, we examine the frequency dependence of the excitation and absorption rates. Figures~\ref{gammaexs} and \ref{gammaabs} illustrate {the corresponding normalized rates $\Gamma_{\mathrm{exc}}/\mathcal{R}=\mathrm{P}_{\mathrm{exc}}$ and $\Gamma_{\mathrm{abs}}/\mathcal{R}=\mathrm{P}_{\mathrm{abs}}$}. The excitation rate $\Gamma_{\mathrm{exc}}$ exhibits a monotonic decrease as the {scalar-mode} frequency $\omega_{\infty}$ grows, while increasing values of the Lorentz--violating parameter $\chi$ further suppress its magnitude. In contrast, the absorption rate $\Gamma_{\mathrm{abs}}$ rises with $\omega_{\infty}$, and larger values of $\chi$ enhance this growth. {Because the individual probabilities were obtained using the hierarchy $\omega_\infty,\bar{\kappa}\ll\tilde{\omega}$, the numerical illustrations must satisfy the same hierarchy and remain in the perturbative regime $\mathrm{P}_{\mathrm{exc/abs}}\ll1$. We therefore use $M=1$, $\mathfrak{a}=1$, $\tilde{\omega}=20$, and $0\leq\omega_\infty\leq0.5$ in Figs.~\ref{gammaexs} and \ref{gammaabs}.}

{\subsubsection{{Entropy flux and radiative area loss}}

At fixed Lorentz violating parameters}, the {HBAR} entropy {flux yields} the {area coefficient $(k_B/4)(1+\ell)$ for the spacelike vacuum and $k_B/4$ for the lightlike vacuum.}

The {detailed balance ratio gives}
\ie
-\ln\left[ \frac{\Gamma_{\mathrm{exc}}} {\Gamma_{\mathrm{abs}}} \right] = -\ln\left[ e^{-8\pi M{\sqrt{q}\,\omega_\infty}} \right] = 8\pi M{\sqrt{q}\,\omega_\infty}.
\fe

Substituting the density matrix elements $\varrho_{n,n}^{{\mathrm{st}}}$ from Eq.~(\ref{vvcccccxx}) into {\hypersetup{citecolor=red,linkcolor=red}Eq.~(\ref{2.46})} {and using}
\begin{equation}
{ \ln\left(\varrho_{n,n}^{\mathrm{st}}\right) = \ln\left( 1-\frac{\Gamma_{\mathrm{exc}}}{\Gamma_{\mathrm{abs}}} \right) +n\ln\left( \frac{\Gamma_{\mathrm{exc}}}{\Gamma_{\mathrm{abs}}} \right) }
\label{logstationarydistribution}
\end{equation}
{together with $\sum_n\dot{\varrho}_{n,n}=0$ yields}
\begin{equation}
\label{kkjjjjff}
\begin{split}
\dot{S}_\varrho {\approx}& \, \, 8\pi M k_{B}{\sqrt{q}} \sum\limits_{\omega_{\infty}} \left( \sum\limits_n n\dot{\varrho}_{n,n} \right) {\omega_\infty} \\ {=}& \,\, 8\pi M k_{B}{\sqrt{q}} \sum\limits_{{\omega_\infty}} {\omega_\infty} \dot{\bar{n}}_{\omega_{\infty}} {=}\,\, 8\pi M k_{B}{\sqrt{q}} {\dot{E}_{\mathrm{rad}}}.
\end{split}
\end{equation}

In Eq.~(\ref{kkjjjjff}), the quantity
$\dot{\bar{n}}_{\omega_{\infty}}
{=\sum_n n\dot{\varrho}_{n,n}}$
represents the {number flux in the mode $\omega_\infty$}. {{We keep} $k_B$ explicit.} Summing over all modes yields the total {outgoing radiative energy flux,
$\dot{E}_{\mathrm{rad}}
=\sum_{\omega_\infty}\omega_\infty\dot{\bar{n}}_{\omega_\infty}>0$,
where all rates are defined with respect to the normalized time $\bar{t}$}. For the configuration considered here, the corresponding black hole horizon area reads
\begin{equation}
\label{assddddddss}
A_{H} = 4\pi r_h^2 {=} 16\pi M^2.
\end{equation}
Differentiating both sides of Eq.~(\ref{assddddddss}) with respect to {the normalized time $\bar{t}$} yields the corresponding evolution equation
\begin{equation}
\dot{A}_{H} = 32\pi M\dot{M}.
\end{equation}

{{\hypersetup{citecolor=red,linkcolor=red}Using the established canonical energy in Eq.~(\ref{canonicalmass})}, {since} $\dot E_{\mathrm{rad}}$ is the energy flux conjugate to $\bar{t}$, energy conservation must instead be imposed on $\mathcal{M}_\infty$. For fixed $\ell$ and $\beta$, imposing $\left.\dot{\mathcal{M}}_\infty^{(\mathrm{S,L})}\right|_{\mathrm{rad}}=-\dot E_{\mathrm{rad}}$ gives}
\begin{equation}
{
\begin{aligned}
\left.\dot M\right|_{\mathrm{rad}}^{(\mathrm{S})}
&=-\frac{\sqrt{q_{\mathrm{S}}}}{1+\ell}\,\dot E_{\mathrm{rad}},
\\
\left.\dot M\right|_{\mathrm{rad}}^{(\mathrm{L})}
&=-\sqrt{q_{\mathrm{L}}}\,\dot E_{\mathrm{rad}}.
\end{aligned}}
\label{canonicalmassloss}
\end{equation}
{If the Lorentz--violating parameters are allowed to vary, $\dot{\mathcal{M}}_\infty$ contains additional work terms; therefore, the area relations below apply only when $\ell$ and $\beta$ are independent of $\bar t$.}

We introduce the contribution to the temporal variation of the black hole horizon area that arises specifically from emission, which we denote as
\begin{equation}
{
\begin{aligned}
\dot{A}_p^{(\mathrm{S})} &\equiv -\left.\dot{A}_H\right|_{\mathrm{rad}}^{(\mathrm{S})} = 32\pi M \frac{\sqrt{q_{\mathrm{S}}}}{1+\ell} \dot{E}_{\mathrm{rad}} >0, \\ \dot{A}_p^{(\mathrm{L})} &\equiv -\left.\dot{A}_H\right|_{\mathrm{rad}}^{(\mathrm{L})} = 32\pi M \sqrt{q_{\mathrm{L}}} \dot{E}_{\mathrm{rad}} >0.
\end{aligned}}
\label{radiativearealoss}
\end{equation}
{Accordingly, $A_p^{(\mathrm{S})}$ and $A_p^{(\mathrm{L})}$ denote the positive area--equivalent losses associated with the emitted radiation; the corresponding contributions to the actual horizon--area variations are $-\dot{A}_p^{(\mathrm{S})}$ and $-\dot{A}_p^{(\mathrm{L})}$, respectively.} This interpretation becomes clearer when the physical sequence of events is examined. An atom emits HBAR radiation while it is still outside the event horizon, prior to {transferring its remaining asymptotically measured energy to the black hole}. As a result, the entropy variation associated with the emitted HBAR radiation can be temporally disentangled from the entropy increase driven by the subsequent incorporation of the atom into the black hole. This separation allows the two effects to be treated as distinct contributions to the total entropy balance.

The resulting expression for $\dot{S}_\varrho$ can thus be expressed explicitly as a function of $A_p$ in the form
\begin{equation}
{ \begin{aligned} \dot{S}_\varrho^{(\mathrm{S})} &\approx \frac{k_B}{4}(1+\ell)\dot{A}_p^{(\mathrm{S})} = \frac{\mathrm{d}}{\mathrm{d}\bar{t}} \left[ \frac{k_B}{4}(1+\ell)A_p^{(\mathrm{S})} \right], \\ \dot{S}_\varrho^{(\mathrm{L})} &\approx \frac{k_B}{4}\dot{A}_p^{(\mathrm{L})} = \frac{\mathrm{d}}{\mathrm{d}\bar{t}} \left[ \frac{k_B}{4}A_p^{(\mathrm{L})} \right].
\end{aligned}}
\label{correctedhbararealaw}
\end{equation}
{In this manner, the factor $\sqrt{q}$ appearing in the spectral entropy-production rate in Eq.~(\ref{kkjjjjff}) does not survive as a universal multiplicative correction to the HBAR area relation. It is canceled by the conversion between the metric parameter $M$ and the asymptotically canonical mass $\mathcal{M}_\infty$. The residual correction is {vacuum} dependent: it is $1+\ell$ for the spacelike VEV and unity for the lightlike VEV.}

{In other words, the corrected HBAR area relation reproduces the same {vacuum} dependent prefactor as the {established} thermodynamic entropy {\hypersetup{citecolor=red,linkcolor=red}in Eq.~(\ref{branchthermodynamicentropy})}. For the spacelike {configuration}, this coefficient must be distinguished from the Wald coefficient, $S_W^{(\mathrm{S})}=(1+\ell/2)k_BA_H/4$, whereas for the lightlike {configuration} $S_W^{(\mathrm{L})}=S_{\mathrm{th}}^{(\mathrm{L})}=k_BA_H/4$. The HBAR result interestingly mirrors the thermodynamic--Wald mismatch of the spacelike solution and its absence in the lightlike solution.}

{A further consequence is that two spacelike and lightlike configurations may have the same value of $q=1+\chi$ and then the same metric and the same detailed-balance factor in Eq.~(\ref{ratebalance}), while still possessing different HBAR area coefficients. The difference arises because the asymptotically canonical energy retains information about the underlying VEV through $\ell_1$, which is not encoded by the metric combination $q$ alone. In other words, this surprisingly shows a way in which the HBAR area relation distinguishes the two {configurations} of the theory even when their geometries coincide. In the Schwarzschild limit, $\ell\rightarrow0$ and $q\rightarrow1$, both cases reduce to $\dot S_\varrho\approx(k_B/4)\dot A_p$, as we should naturally expect.}

At this stage, it is worth noting that, within the HBAR framework, several Lorentz--violating black hole configurations have already been examined, including bumblebee gravity \cite{Rahaman:2025grm}, Kalb--Ramond gravity \cite{Rahaman:2025mrr}, its extension incorporating a global monopole sector \cite{Pantig:2025okn}, as well as scenarios inspired by the generalized uncertainty principle \cite{Ovgun:2025isv}.



\section{{Conclusion}}\label{Sec:Conclusion}

{We examined the near horizon} quantum {physics} of {the adopted} bumblebee black hole {through scalar field correlations}, the {excitation} of a freely falling atom, and {HBAR} entropy. The {correlation measures} depended on the geometry through $\chi=\alpha\ell$. {Spacelike} and lightlike {vacuum} configurations with {equal} $\chi$ produced identical logarithmic negativity and mutual information. At fixed $\beta$ and $\ell$, {their different geometric deformations produced} the relative deviations {\hypersetup{citecolor=red,linkcolor=red}shown in Fig.~\ref{fig3}. Positive $\chi$ partially counteracted entanglement degradation, with} the {largest differences occurring at low packet frequencies}.

{Using} the {asymptotically normalized time coordinate}, we {obtained} the atomic {excitation probability} in terms of {$\omega_\infty$} and {$\bar\kappa=1/(4M\sqrt q)$}. {At} leading near horizon order and under {$\omega_\infty,\bar\kappa\ll\tilde\omega$, its local thermal factor agreed with} the {flat spacetime acceleration radiation expression. This result supported} the {local} equivalence principle {interpretation within} the {approximations adopted}.

{The} HBAR {calculation gave $\Gamma_{\mathrm{exc}}/\Gamma_{\mathrm{abs}}=\exp[-8\pi M\sqrt q\,\omega_\infty]$ and $\dot S_\varrho\simeq8\pi M k_B\sqrt q\,\dot E_{\mathrm{rad}}$}. {Positive} $\chi$ suppressed the normalized excitation rate and enhanced the normalized absorption rate {in} the perturbative regime {illustrated}. {Using} the {established canonical energy $\mathcal M_\infty=(1+\ell_1)M/\sqrt q$ converted} the spectral entropy {flux into $\dot S_\varrho^{(\mathrm S)}\simeq(k_B/4)(1+\ell)\dot A_p^{(\mathrm S)}$ and $\dot S_\varrho^{(\mathrm L)}\simeq(k_B/4)\dot A_p^{(\mathrm L)}$. Here $\dot A_p>0$ denoted} the {radiative} area {loss}. {The area coefficients distinguished} the {two vacuum configurations even when their geometries and detailed balance factors agreed.

These coefficients reproduced} the {known thermodynamic entropy factors}. The HBAR area relation {thus reflected} the {established thermodynamic--Wald entropy difference for the spacelike vacuum} and the {agreement} of the two entropies for the lightlike {vacuum}. {The derivation held} $\ell$ and $\beta$ {fixed; varying these} parameters {introduced} additional work terms in the {canonical energy balance}. {Both cases recovered $\dot S_\varrho\simeq(k_B/4)\dot A_p$ in} the Schwarzschild limit.

{ It would also be natural} to extend the present framework to more general scenarios, including spacetimes with multiple horizons and other nontrivial background geometries, where richer near--horizon structures could generate new patterns of quantum correlations and information transport \cite{Wu:2023vis,Bao:2025dzj,Wu:2024nfp,Liu:2025ctf,Li:2024pwo,Du:2024rtc}. In such settings, Lorentz symmetry breaking may lead to multi--directional effects, in which several preferred spacetime directions affect quantum probes in a nontrivial way \cite{Liu:2025lui}. Furthermore, investigating possible links between near--horizon quantum--information measures and horizon--scale observables, such as black hole shadows \cite{Perlick:2021aok,Cunha:2018acu,Liu:2025wwq,Liu:2024soc,Liu:2024iec,Liu:2024lbi,Errehymy:2025ada}, may help connect microscopic vacuum properties with observable strong--field signatures.

\section*{Acknowledgments}
\hspace{0.5cm} A.A.A.F. is supported by Conselho Nacional de Desenvolvimento Cient\'{\i}fico e Tecnol\'{o}gico (CNPq) -- [150223/2025-0]. A.A.A.F. thanks A. Övgun for drawing attention to the topic of HBAR entropy and for sharing the main related references. 
W. Liu acknowledges support from the China Postdoctoral Science Foundation (Grant No. 2025M783393) and the National Natural Science Foundation of China (Grant No. 12547147).

\section*{Data Availability Statement}

Data Availability Statement: {The numerical data underlying the results and figures presented in this manuscript, together with the computational codes used to generate them, are available from the corresponding author upon reasonable request.}

\bibliographystyle{ieeetr}
\bibliography{main}

\begin{thebibliography}{100}

\bibitem{Hu:2018mni}
M.-L. Hu, X.~Hu, J.~Wang, Y.~Peng, Y.-R. Zhang, and H.~Fan, ``{Quantum
  coherence and geometric quantum discord},'' {\em Phys. Rept.}, vol.~762-764,
  pp.~1--100, 2018.

\bibitem{Boschi:1997dg}
D.~Boschi, S.~Branca, F.~De~Martini, L.~Hardy, and S.~Popescu, ``{Experimental
  realization of teleporting an unknown pure quantum state via dual classical
  and Einstein-Podolski-Rosen channels},'' {\em Phys. Rev. Lett.}, vol.~80,
  pp.~1121--1125, 1998.

\bibitem{Pan:2000vtd}
J.-W. Pan, C.~Simon, {\v{C}}.~Brukner, and A.~Zeilinger, ``{Entanglement
  purification for quantum communication},'' {\em Nature}, vol.~410, no.~6832,
  pp.~1067--1070, 2001.

\bibitem{Aziz:2025ypo}
J.~Aziz and R.~Howl, ``{Classical theories of gravity produce entanglement},''
  {\em Nature}, vol.~646, no.~8086, pp.~813--817, 2025.

\bibitem{Unruh:1976db}
W.~G. Unruh, ``{Notes on black hole evaporation},'' {\em Phys. Rev. D},
  vol.~14, p.~870, 1976.

\bibitem{Adesso:2007wi}
G.~Adesso, I.~Fuentes-Schuller, and M.~Ericsson, ``{Continuous variable
  entanglement sharing in non-inertial frames},'' {\em Phys. Rev. A}, vol.~76,
  p.~062112, 2007.

\bibitem{Barros:2023kor}
P.~H.~M. Barros, I.~G. da~Paz, O.~P. d.~S. Neto, and H.~A.~S. Costa,
  ``{Robustness of Wave{\textendash}Particle Duality under Unruh Effect},''
  {\em Entropy}, vol.~26, no.~1, p.~1, 2024.

\bibitem{Barros:2024wdv}
P.~H.~M. Barros and H.~A.~S. Costa, ``{Detecting gravitational waves via
  coherence degradation induced by the Unruh effect},'' {\em Eur. Phys. J. C},
  vol.~84, no.~12, p.~1261, 2024.

\bibitem{Kollas:2022wgj}
N.~K. Kollas, D.~Moustos, and M.~R. Mu{\~n}oz, ``{Cohering and decohering power
  of massive scalar fields under instantaneous interactions},'' {\em Phys. Rev.
  A}, vol.~107, no.~2, p.~022420, 2023.

\bibitem{Barros:2025arh}
P.~H.~M. Barros, P.~R.~S. Carvalho, and H.~A.~S. Costa, ``{On the information
  behavior from quadratically coupled accelerated detectors},'' {\em Eur. Phys.
  J. C}, vol.~85, no.~8, p.~868, 2025.

\bibitem{Barros:2025din}
P.~H.~M. Barros, F.~C.~E. Lima, C.~A.~S. Almeida, and H.~A.~S. Costa,
  ``{Mitigating the information degradation in a massive Unruh-DeWitt
  theory},'' {\em JHEP}, vol.~04, p.~165, 2025.

\bibitem{Fuentes-Schuller:2004iaz}
I.~Fuentes-Schuller and R.~B. Mann, ``{Alice falls into a black hole:
  Entanglement in non-inertial frames},'' {\em Phys. Rev. Lett.}, vol.~95,
  p.~120404, 2005.

\bibitem{Fuentes:2010dt}
I.~Fuentes, R.~B. Mann, E.~Martin-Martinez, and S.~Moradi, ``{Entanglement of
  Dirac fields in an expanding spacetime},'' {\em Phys. Rev. D}, vol.~82,
  p.~045030, 2010.

\bibitem{Alsing:2006cj}
P.~M. Alsing, I.~Fuentes-Schuller, R.~B. Mann, and T.~E. Tessier,
  ``{Entanglement of Dirac fields in non-inertial frames},'' {\em Phys. Rev.
  A}, vol.~74, p.~032326, 2006.

\bibitem{Martin-Martinez:2010yva}
E.~Martin-Martinez, L.~J. Garay, and J.~Leon, ``{Unveiling quantum entanglement
  degradation near a Schwarzschild black hole},'' {\em Phys. Rev. D}, vol.~82,
  p.~064006, 2010.

\bibitem{Liu:2024wpa}
W.~Liu, C.~Wen, and J.~Wang, ``{Lorentz violation alleviates gravitationally
  induced entanglement degradation},'' {\em JHEP}, vol.~01, p.~184, 2025.

\bibitem{Liu:2015oat}
X.~Liu, Z.~Tian, J.~Wang, and J.~Jing, ``{Protecting quantum coherence of
  two-level atoms from vacuum fluctuations of electromagnetic field},'' {\em
  Annals Phys.}, vol.~366, pp.~102--112, 2016.

\bibitem{Wu:2023sye}
S.-M. Wu, X.-W. Fan, R.-D. Wang, H.-Y. Wu, X.-L. Huang, and H.-S. Zeng, ``{Does
  Hawking effect always degrade fidelity of quantum teleportation in
  Schwarzschild spacetime?},'' {\em JHEP}, vol.~11, p.~232, 2023.

\bibitem{Liu:2023lok}
X.~Liu, Z.~Tian, and J.~Jing, ``{Entanglement dynamics in
  {\ensuremath{\kappa}}-deformed spacetime},'' {\em Sci. China Phys. Mech.
  Astron.}, vol.~67, no.~10, p.~100411, 2024.

\bibitem{Sen:2023sfb}
S.~Sen, A.~Mukherjee, and S.~Gangopadhyay, ``{Entanglement degradation as a
  tool to detect signatures of modified gravity},'' {\em Phys. Rev. D},
  vol.~109, no.~4, p.~046012, 2024.

\bibitem{Li:2022pwa}
W.-M. Li, S.-M. Wu, H.-S. Zeng, and X.-L. Huang, ``{Bosonic and fermionic
  coherence of N-partite states in the background of a dilaton black hole},''
  {\em Quant. Inf. Proc.}, vol.~21, no.~10, p.~362, 2022.

\bibitem{Wu:2022lmc}
S.-M. Wu, C.-X. Wang, D.-D. Liu, X.-L. Huang, and H.-S. Zeng, ``{Would quantum
  coherence be increased by curvature effect in de Sitter space?},'' {\em
  JHEP}, vol.~02, p.~115, 2023.

\bibitem{Wu:2022xwy}
S.-M. Wu and H.-S. Zeng, ``{Genuine tripartite nonlocality and entanglement in
  curved spacetime},'' {\em Eur. Phys. J. C}, vol.~82, no.~1, p.~4, 2022.

\bibitem{Babakan:2024abb}
M.~Babakan, A.~Kashef, and L.~Memarzadeh, ``{Entanglement degradation under
  local dissipative Landau-Zener noise},'' {\em Phys. Rev. A}, vol.~110, no.~1,
  p.~012456, 2024.

\bibitem{Wu:2023spa}
S.-M. Wu, X.-W. Teng, J.-X. Li, S.-H. Li, T.-H. Liu, and J.-C. Wang,
  ``{Genuinely accessible and inaccessible entanglement in Schwarzschild black
  hole},'' {\em Phys. Lett. B}, vol.~848, p.~138334, 2024.

\bibitem{Wu:2024qhd}
S.-M. Wu, R.-D. Wang, X.-L. Huang, and Z.~Wang, ``{Does gravitational wave
  assist vacuum steering and Bell nonlocality?},'' {\em JHEP}, vol.~07, p.~155,
  2024.

\bibitem{AraujoFilho:2025rwr}
A.~A. Ara{\'u}jo~Filho, ``{Particle production induced by a Lorentzian
  non-commutative spacetime},'' {\em Annals Phys.}, vol.~481, p.~170167, 2025.

\bibitem{AraujoFilho:2025hkm}
A.~A. Ara{\'u}jo~Filho, ``{How does non-metricity affect particle creation and
  evaporation in bumblebee gravity?},'' {\em JCAP}, vol.~06, p.~026, 2025.

\bibitem{Wu:2025euf}
S.-M. Wu, X.-W. Teng, W.-M. Li, Y.-X. Wang, and J.~Lu, ``{Nonseparability of
  multipartite systems in dilaton black~hole},'' {\em JCAP}, vol.~09, p.~030,
  2025.

\bibitem{Li:2025jlu}
W.-M. Li, J.~Lu, and S.-M. Wu, ``{Multiqubit coherence of mixed states near
  event horizon},'' 5 2025.

\bibitem{Wu:2024urq}
S.-M. Wu and S.-H. Li, ``{Do maximally entangled states always have an
  advantage over non-maximally entangled states in Schwarzschild black
  hole?},'' {\em Eur. Phys. J. C}, vol.~85, no.~6, p.~614, 2025.

\bibitem{Huang:2024vyc}
X.~Huang, H.~Wu, and S.~Wu, ``{Generated genuine tripartite steering and its
  monogamy in the background of a Kerr-Newman black hole},'' {\em Chin. Phys.
  C}, vol.~48, no.~11, p.~115106, 2024.

\bibitem{Liu:2025hcx}
X.~Liu, W.~Liu, and S.-M. Wu, ``{Entanglement degradation of static black holes
  in effective quantum gravity},'' 11 2025.

\bibitem{Wu:2025ncd}
S.-M. Wu, Y.-X. Wang, S.-H. Shang, and W.~Liu, ``{Influence of dark matter on
  quantum entanglement and coherence in curved spacetime},'' 7 2025.

\bibitem{Du:2024ihk}
M.-M. Du, H.-W. Li, S.-T. Shen, X.-J. Yan, X.-Y. Li, L.~Zhou, W.~Zhong, and
  Y.-B. Sheng, ``{Maximal steered coherence in the background of Schwarzschild
  space-time},'' {\em Eur. Phys. J. C}, vol.~84, no.~5, p.~450, 2024.

\bibitem{Liu:2026vvp}
S.-Y. Liu, X.-Z. Song, X.-Y. Yu, W.~Liu, X.-L. Huang, and S.-M. Wu,
  ``{Asymmetric quantum steering harvested near a Lorentz-violating BTZ black
  hole},'' 6 2026.

\bibitem{Louko:2006zv}
J.~Louko and A.~Satz, ``{How often does the Unruh-DeWitt detector click?
  Regularisation by a spatial profile},'' {\em Class. Quant. Grav.}, vol.~23,
  pp.~6321--6344, 2006.

\bibitem{Du:2025jiv}
M.-M. Du, H.-W. Li, S.-T. Shen, X.-J. Yan, X.-Y. Li, L.~Zhou, W.~Zhong, and
  Y.-B. Sheng, ``{Entropic Uncertainty Relations with Quantum Memory in
  Accelerated Frames via Unruh-DeWitt Detectors},'' 12 2025.

\bibitem{Li:2024wtc}
H.-W. Li, Y.-H. Fan, S.-T. Shen, X.-J. Yan, X.-Y. Li, W.~Zhong, Y.-B. Sheng,
  L.~Zhou, and M.-M. Du, ``{Maximal steered coherence in accelerating
  Unruh{\textendash}DeWitt detectors},'' {\em Eur. Phys. J. C}, vol.~84,
  no.~12, p.~1241, 2024.

\bibitem{Wu:2025qqu}
S.-M. Wu, Y.-X. Wang, and W.~Liu, ``{Entangled Unruh{\textendash}DeWitt
  detectors amplify quantum coherence},'' {\em Eur. Phys. J. C}, vol.~85,
  no.~10, p.~1095, 2025.

\bibitem{Henderson:2017yuv}
L.~J. Henderson, R.~A. Hennigar, R.~B. Mann, A.~R.~H. Smith, and J.~Zhang,
  ``{Harvesting Entanglement from the Black Hole Vacuum},'' {\em Class. Quant.
  Grav.}, vol.~35, no.~21, p.~21LT02, 2018.

\bibitem{Tjoa:2020eqh}
E.~Tjoa and R.~B. Mann, ``{Harvesting correlations in Schwarzschild and
  collapsing shell spacetimes},'' {\em JHEP}, vol.~08, p.~155, 2020.

\bibitem{Cong:2018vqx}
W.~Cong, E.~Tjoa, and R.~B. Mann, ``{Entanglement Harvesting with Moving
  Mirrors},'' {\em JHEP}, vol.~06, p.~021, 2019.
\newblock [Erratum: JHEP 07, 051 (2019)].

\bibitem{Foo:2020dzt}
J.~Foo, S.~Onoe, R.~B. Mann, and M.~Zych, ``{Thermality, causality, and the
  quantum-controlled Unruh{\textendash}deWitt detector},'' {\em Phys. Rev.
  Res.}, vol.~3, no.~4, p.~043056, 2021.

\bibitem{Zhang:2020xvo}
J.~Zhang and H.~Yu, ``{Entanglement harvesting for Unruh-DeWitt detectors in
  circular motion},'' {\em Phys. Rev. D}, vol.~102, no.~6, p.~065013, 2020.

\bibitem{Gallock-Yoshimura:2021yok}
K.~Gallock-Yoshimura, E.~Tjoa, and R.~B. Mann, ``{Harvesting entanglement with
  detectors freely falling into a black hole},'' {\em Phys. Rev. D}, vol.~104,
  no.~2, p.~025001, 2021.

\bibitem{Bueley:2022ple}
K.~Bueley, L.~Huang, K.~Gallock-Yoshimura, and R.~B. Mann, ``{Harvesting mutual
  information from BTZ black hole spacetime},'' {\em Phys. Rev. D}, vol.~106,
  no.~2, p.~025010, 2022.

\bibitem{Zhou:2021nyv}
Y.~Zhou, J.~Hu, and H.~Yu, ``{Entanglement dynamics for Unruh-DeWitt detectors
  interacting with massive scalar fields: the Unruh and anti-Unruh effects},''
  {\em JHEP}, vol.~09, p.~088, 2021.

\bibitem{Li:2025bzd}
S.-H. Li, S.-H. Shang, and S.-M. Wu, ``{Does acceleration always degrade
  quantum entanglement for tetrapartite Unruh-DeWitt detectors?},'' {\em JHEP},
  vol.~05, p.~214, 2025.

\bibitem{Liu:2025zts}
Z.~Liu, W.~Liu, X.~Liu, and J.~Wang, ``{Wormhole-Induced correlation: A Link
  Between Two Universes},'' 10 2025.

\bibitem{Tang:2025mtc}
Y.~Tang, W.~Liu, Z.~Liu, and J.~Wang, ``{Can the latent signatures of quantum
  superposition be detected through correlation harvesting?},'' 8 2025.

\bibitem{Liu:2025bpp}
X.~Liu, W.~Liu, Z.~Liu, and J.~Wang, ``{Harvesting correlations from BTZ black
  hole coupled to a Lorentz-violating vector field},'' {\em JHEP}, vol.~08,
  p.~094, 2025.

\bibitem{colladay1997cpt}
D.~Colladay and V.~A. Kosteleck{\`y}, ``Cpt violation and the standard model,''
  {\em Physical Review D}, vol.~55, no.~11, p.~6760, 1997.

\bibitem{kostelecky1989spontaneous}
V.~A. Kosteleck{\`y} and S.~Samuel, ``Spontaneous breaking of lorentz symmetry
  in string theory,'' {\em Physical Review D}, vol.~39, no.~2, p.~683, 1989.

\bibitem{kostelecky2004gravity}
V.~A. Kosteleck{\`y}, ``Gravity, lorentz violation, and the standard model,''
  {\em Physical Review D}, vol.~69, no.~10, p.~105009, 2004.

\bibitem{kostelecky2011data}
V.~A. Kosteleck{\`y} and N.~Russell, ``Data tables for lorentz and cpt
  violation,'' {\em Reviews of Modern Physics}, vol.~83, no.~1, pp.~11--31,
  2011.

\bibitem{kostelecky1999constraints}
V.~A. Kosteleck{\`y} and C.~D. Lane, ``Constraints on lorentz violation from
  clock-comparison experiments,'' {\em Physical Review D}, vol.~60, no.~11,
  p.~116010, 1999.

\bibitem{Bluhm:2023kph}
R.~Bluhm and Y.~Zhi, ``{Spontaneous and Explicit Spacetime Symmetry Breaking in
  Einstein{\textendash}Cartan Theory with Background Fields},'' {\em Symmetry},
  vol.~16, no.~1, p.~25, 2024.

\bibitem{Maluf:2013nva}
R.~V. Maluf, V.~Santos, W.~T. Cruz, and C.~A.~S. Almeida, ``{Matter-gravity
  scattering in the presence of spontaneous Lorentz violation},'' {\em Phys.
  Rev. D}, vol.~88, no.~2, p.~025005, 2013.

\bibitem{bluhm2005spontaneous}
R.~Bluhm and V.~A. Kosteleck{\`y}, ``Spontaneous lorentz violation,
  nambu-goldstone modes, and gravity,'' {\em Physical Review D—Particles,
  Fields, Gravitation, and Cosmology}, vol.~71, no.~6, p.~065008, 2005.

\bibitem{Bluhm:2019ato}
R.~Bluhm, H.~Bossi, and Y.~Wen, ``{Gravity with explicit spacetime symmetry
  breaking and the Standard-Model Extension},'' {\em Phys. Rev. D}, vol.~100,
  no.~8, p.~084022, 2019.

\bibitem{Maluf:2014dpa}
R.~V. Maluf, C.~A.~S. Almeida, R.~Casana, and M.~M. Ferreira, Jr.,
  ``{Einstein-Hilbert graviton modes modified by the Lorentz-violating
  bumblebee Field},'' {\em Phys. Rev. D}, vol.~90, no.~2, p.~025007, 2014.

\bibitem{bluhm2008spontaneous}
R.~Bluhm, S.-H. Fung, and V.~A. Kosteleck{\`y}, ``Spontaneous lorentz and
  diffeomorphism violation, massive modes, and gravity,'' {\em Physical Review
  D—Particles, Fields, Gravitation, and Cosmology}, vol.~77, no.~6,
  p.~065020, 2008.

\bibitem{Liu:2025lwj}
W.~Liu, H.~Huang, D.~Wu, and J.~Wang, ``{Lorentz violation signatures in the
  low-energy sector of Ho{\v{r}}ava gravity from black hole shadow
  observations},'' {\em Phys. Lett. B}, vol.~868, p.~139812, 2025.

\bibitem{Casana:2017jkc}
R.~Casana, A.~Cavalcante, F.~P. Poulis, and E.~B. Santos, ``{Exact
  Schwarzschild-like solution in a bumblebee gravity model},'' {\em Phys. Rev.
  D}, vol.~97, no.~10, p.~104001, 2018.

\bibitem{Liu:2024axg}
J.-Z. Liu, W.-D. Guo, S.-W. Wei, and Y.-X. Liu, ``{Charged spherically
  symmetric and slowly rotating charged black hole solutions in bumblebee
  gravity},'' {\em Eur. Phys. J. C}, vol.~85, no.~2, p.~145, 2025.

\bibitem{Chen:2025ypx}
Y.-Q. Chen and H.-S. Liu, ``{Taub-NUT-like black holes in Einstein-bumblebee
  gravity},'' {\em Phys. Rev. D}, vol.~112, no.~8, p.~084040, 2025.

\bibitem{Liu:2022dcn}
W.~Liu, X.~Fang, J.~Jing, and J.~Wang, ``{QNMs of slowly rotating
  Einstein{\textendash}Bumblebee black hole},'' {\em Eur. Phys. J. C}, vol.~83,
  no.~1, p.~83, 2023.

\bibitem{Liu:2024oeq}
W.~Liu, X.~Fang, J.~Jing, and J.~Wang, ``{Lorentz violation induces
  isospectrality breaking in Einstein-bumblebee gravity theory},'' {\em Sci.
  China Phys. Mech. Astron.}, vol.~67, no.~8, p.~280413, 2024.

\bibitem{Deng:2025uvp}
W.~Deng, W.~Liu, F.~Long, K.~Xiao, and J.~Jing, ``{Quasinormal modes of a
  massive scalar field in slowly rotating Einstein-Bumblebee black holes},''
  {\em JCAP}, vol.~11, p.~028, 2025.

\bibitem{Tang:2025eew}
Y.~Tang, W.~Liu, and J.~Wang, ``{Observational signature of Lorentz violation
  in acceleration radiation},'' {\em Eur. Phys. J. C}, vol.~85, no.~10,
  p.~1108, 2025.

\bibitem{Sekhmani:2025zen}
Y.~Sekhmani, W.~Liu, W.~Deng, and K.~Boshkayev, ``{Quasinormal Modes of Massive
  Scalar Perturbations in Slow-Rotation Bumblebee Black Holes with Traceless
  Conformal Electrodynamics},'' 10 2025.

\bibitem{Li:2025itp}
B.-R. Li, J.-Z. Liu, W.-D. Guo, and Y.-X. Liu, ``{Quasinormal modes of a
  charged spherically symmetric black hole in bumblebee gravity},'' 10 2025.

\bibitem{Lai:2025nyo}
X.-B. Lai, Y.-Q. Dong, Y.-Z. Fan, and Y.-X. Liu, ``{Stability Analysis of
  Cosmological Perturbations in the Bumblebee Model: Parameter Constraints and
  Gravitational Waves},'' 9 2025.

\bibitem{Yang:2023wtu}
K.~Yang, Y.-Z. Chen, Z.-Q. Duan, and J.-Y. Zhao, ``{Static and spherically
  symmetric black holes in gravity with a background Kalb-Ramond field},'' {\em
  Phys. Rev. D}, vol.~108, no.~12, p.~124004, 2023.

\bibitem{Liu:2024oas}
W.~Liu, D.~Wu, and J.~Wang, ``{Static neutral black holes in Kalb-Ramond
  gravity},'' {\em JCAP}, vol.~09, p.~017, 2024.

\bibitem{Liu:2024lve}
W.~Liu, D.~Wu, and J.~Wang, ``{Shadow of slowly rotating Kalb-Ramond black
  holes},'' {\em JCAP}, vol.~05, p.~017, 2025.

\bibitem{Liu:2025fxj}
J.-Z. Liu, S.-P. Wu, S.-W. Wei, and Y.-X. Liu, ``{Exact black hole solutions in
  gravity with a background Kalb-Ramond field},'' {\em JCAP}, vol.~11, p.~056,
  2025.

\bibitem{Yu:2025odj}
Z.-X. Yu, H.-D. Lyu, M.~Huhe, and S.~Li, ``{Revisiting black holes and their
  thermodynamics in Einstein-Kalb-Ramond gravity},'' 11 2025.

\bibitem{Deng:2025atg}
W.~Deng, W.~Liu, K.~Xiao, and J.~Jing, ``{Quasinormal modes of scalar,
  electromagnetic, and gravitational perturbations in slowly rotating
  Kalb-Ramond black holes},'' 11 2025.

\bibitem{Gu:2025lyz}
Y.-T. Gu, W.-D. Guo, and Y.-X. Liu, ``{Quasinormal modes of an electrically
  charged Kalb-Ramond black hole},'' 9 2025.

\bibitem{Guo:2023nkd}
W.-D. Guo, Q.~Tan, and Y.-X. Liu, ``{Quasinormal modes and greybody factor of a
  Lorentz-violating black hole},'' {\em JCAP}, vol.~07, p.~008, 2024.

\bibitem{Xia:2025hwt}
Z.-W. Xia, S.~Long, H.~Gong, Q.~Pan, and J.~Jing, ``{Scalar perturbation around
  a rotating Kalb-Ramond BTZ black hole},'' 11 2025.

\bibitem{Liu:2026khl}
W.~Liu, D.~Wu, X.~Fang, Y.-X. Liu, and J.~Wang, ``{Macroscopic Optical
  Nonreciprocity: A Black Hole as an Optical Diode},'' 4 2026.

\bibitem{Liu:2026cxs}
H.-F. Liu, W.~Liu, Y.-X. Liu, Q.~Su, and D.-f. Zeng, ``{Gravitational-Bumblebee
  perturbations: Exact decoupling and isospectrality},'' 5 2026.

\bibitem{Bertolami:2005bh}
O.~Bertolami and J.~Paramos, ``{The Flight of the bumblebee: Vacuum solutions
  of a gravity model with vector-induced spontaneous Lorentz symmetry
  breaking},'' {\em Phys. Rev. D}, vol.~72, p.~044001, 2005.

\bibitem{AraujoFilho:2024ctw}
A.~A. Ara{\'u}jo~Filho, ``{Particle creation and evaporation in Kalb-Ramond
  gravity},'' {\em JCAP}, vol.~04, p.~076, 2025.

\bibitem{Liang:2022hxd}
D.~Liang, R.~Xu, X.~Lu, and L.~Shao, ``{Polarizations of gravitational waves in
  the bumblebee gravity model},'' {\em Phys. Rev. D}, vol.~106, no.~12,
  p.~124019, 2022.

\bibitem{amarilo2024gravitational}
K.~M. Amarilo, M.~B. Ferreira~Filho, A.~A. Ara{\'u}jo~Filho, and J.~A. A.~S.
  Reis, ``Gravitational waves effects in a lorentz--violating scenario,'' {\em
  Physics Letters B}, vol.~855, p.~138785, 2024.

\bibitem{Maluf:2020kgf}
R.~V. Maluf and J.~C.~S. Neves, ``{Black holes with a cosmological constant in
  bumblebee gravity},'' {\em Phys. Rev. D}, vol.~103, no.~4, p.~044002, 2021.

\bibitem{Uniyal:2022xnq}
A.~Uniyal, S.~Kanzi, and {\.I}.~Sakall{\i}, ``{Some observable physical
  properties of the higher dimensional dS/AdS black holes in Einstein-bumblebee
  gravity theory},'' {\em Eur. Phys. J. C}, vol.~83, no.~7, p.~668, 2023.

\bibitem{Filho:2022yrk}
A.~A.~A. Filho, J.~R. Nascimento, A.~Y. Petrov, and P.~J. Porf{\'\i}rio,
  ``{Vacuum solution within a metric-affine bumblebee gravity},'' {\em Phys.
  Rev. D}, vol.~108, no.~8, p.~085010, 2023.

\bibitem{AraujoFilho:2024ykw}
A.~A. Ara{\'u}jo~Filho, J.~R. Nascimento, A.~Y. Petrov, and P.~J.
  Porf{\'\i}rio, ``{An exact stationary axisymmetric vacuum solution within a
  metric-affine bumblebee gravity},'' {\em JCAP}, vol.~07, p.~004, 2024.

\bibitem{AraujoFilho:2025rvn}
A.~A. Ara{\'u}jo~Filho, N.~Heidari, I.~P. Lobo, Y.~Shi, and F.~S.~N. Lobo,
  ``{The Flight of the Bumblebee in a Non-Commutative Geometry: A New Black
  Hole Solution},'' 9 2025.

\bibitem{AraujoFilho:2025jcu}
A.~A. Ara{\'u}jo~Filho, N.~Heidari, and I.~P. Lobo, ``{A non-commutative
  Kalb-Ramond black hole},'' {\em JCAP}, vol.~09, p.~076, 2025.

\bibitem{Magalhaes:2025nql}
R.~B. Magalh{\~a}es, L.~A. Lessa, and M.~M. Ferreira, ``{Wormholes in
  Lorentz-violating gravity},'' {\em Eur. Phys. J. Plus}, vol.~141, no.~3,
  p.~269, 2026.

\bibitem{AraujoFilho:2024iox}
A.~A. Ara{\'u}jo~Filho, J.~A. A.~S. Reis, and A.~{\"O}vg{\"u}n, ``{Modified
  particle dynamics and thermodynamics in a traversable wormhole in bumblebee
  gravity},'' {\em Eur. Phys. J. C}, vol.~85, no.~1, p.~83, 2025.

\bibitem{Ovgun:2018xys}
A.~{\"O}vg{\"u}n, K.~Jusufi, and {\.I}.~Sakall{\i}, ``{Exact traversable
  wormhole solution in bumblebee gravity},'' {\em Phys. Rev. D}, vol.~99,
  no.~2, p.~024042, 2019.

\bibitem{Magalhaes:2025lti}
R.~B. Magalh{\~a}es, L.~A. Lessa, and R.~Casana, ``{Lorentz-violating
  wormholes: The role of the matter coupled to Lorentz-violating fields},'' 7
  2025.

\bibitem{Pereira:2025xnw}
C.~F.~S. Pereira, M.~V. d.~S. Silva, H.~Belich, D.~C.~Rodrigues, J.~C. Fabris,
  and M.~E. Rodrigues, ``{Black-bounce solutions in a k-essence theory under
  the effects of bumblebee gravity},'' {\em Phys. Rev. D}, vol.~111, no.~12,
  p.~124005, 2025.

\bibitem{Shi:2025plr}
Y.~Shi and A.~A. Ara{\'u}jo~Filho, ``Effects of bumblebee gravity on neutrino
  motion,'' {\em Journal of Cosmology and Astroparticle Physics}, vol.~2025,
  no.~11, p.~045, 2025.

\bibitem{Shi:2025ywa}
Y.~Shi and A.~A. Ara{\'u}jo~Filho, ``{The role of non-metricity on neutrino
  behavior in bumblebee gravity},'' {\em Annals Phys.}, vol.~490, p.~170496,
  2026.

\bibitem{Shi:2025rfq}
Y.~Shi and A.~A. Ara{\'u}jo~Filho, ``{Influence of a Kalb-Ramond black hole on
  neutrino behavior},'' {\em JHEP}, vol.~08, p.~028, 2025.

\bibitem{Shi:2025xkd}
Y.~Shi, A.~A. Ara{\'u}jo~Filho, K.~E.~L. de~Farias, V.~B. Bezerra, and A.~R.
  Queiroz, ``{Neutrino oscillations in a Kalb-Ramond black hole background},''
  12 2025.

\bibitem{Khodadi:2021owg}
M.~Khodadi, ``{Black Hole Superradiance in the Presence of Lorentz Symmetry
  Violation},'' {\em Phys. Rev. D}, vol.~103, no.~6, p.~064051, 2021.

\bibitem{Khodadi:2023yiw}
M.~Khodadi, G.~Lambiase, and L.~Mastrototaro, ``{Spontaneous Lorentz symmetry
  breaking effects on GRBs jets arising from neutrino pair annihilation process
  near a black hole},'' {\em Eur. Phys. J. C}, vol.~83, no.~3, p.~239, 2023.

\bibitem{Khodadi:2022mzt}
M.~Khodadi, G.~Lambiase, and A.~Sheykhi, ``{Constraining the Lorentz-violating
  bumblebee vector field with big bang nucleosynthesis and gravitational
  baryogenesis},'' {\em Eur. Phys. J. C}, vol.~83, no.~5, p.~386, 2023.

\bibitem{Khodadi:2022dff}
M.~Khodadi, ``{Magnetic reconnection and energy extraction from a spinning
  black hole with broken Lorentz symmetry},'' {\em Phys. Rev. D}, vol.~105,
  no.~2, p.~023025, 2022.

\bibitem{Zhu:2025fiy}
J.~Zhu and H.~Li, ``{Full classification of static spherical vacuum solutions
  to bumblebee gravity with general VEVs},'' {\em Phys. Lett. B}, vol.~876,
  p.~140396, 2026.

\bibitem{Liu:2025oho}
J.-Z. Liu, S.-P. Wu, S.-W. Wei, and Y.-X. Liu, ``{Exact black hole solutions in
  bumblebee gravity with lightlike or spacelike VEVs},'' {\em Sci. China Phys.
  Mech. Astron.}, vol.~69, no.~7, p.~270411, 2026.

\bibitem{AraujoFilho:2025zaj}
A.~A. Ara{\'u}jo~Filho, N.~Heidari, I.~P. Lobo, and V.~B. Bezerra,
  ``{Gravitational aspects of a new bumblebee black hole},'' 11 2025.

\bibitem{Shi:2025tvu}
Y.~Shi and A.~A. Ara{\'u}jo~Filho, ``{Neutrino oscillations induced by a new
  bumblebee black hole},'' 11 2025.

\bibitem{Kumar:2025bim}
A.~Kumar, S.~U. Islam, and S.~G. Ghosh, ``{Probing Lorentz symmetry violation
  through lensing observables of rotating black holes},'' {\em Phys. Dark
  Univ.}, vol.~52, p.~102307, 2026.

\bibitem{Shi:2025hfe}
Y.~Shi and A.~A. Ara{\'u}jo~Filho, ``{Accretion of matter of a new bumblebee
  black hole},'' {\em Nucl. Phys. B}, vol.~1023, p.~117327, 2026.

\bibitem{Heidari:2025oop}
N.~Heidari and A.~A. Ara{\'u}jo~Filho, ``{Quantum particle production and
  radiative properties of a new bumblebee black hole},'' 12 2025.

\bibitem{Bruschi:2010yva}
D.~E. Bruschi, J.~Louko, E.~Mart\'{\i}n-Mart\'{\i}nez, A.~Dragan, and
  I.~Fuentes, ``The unruh effect in quantum information beyond the single-mode
  approximation,'' {\em Phys. Rev. A}, vol.~82, p.~042332, 2010.

\bibitem{scully2018quantum}
M.~O. Scully, S.~Fulling, D.~M. Lee, D.~N. Page, W.~P. Schleich, and A.~A.
  Svidzinsky, ``Quantum optics approach to radiation from atoms falling into a
  black hole,'' {\em Proceedings of the National Academy of Sciences},
  vol.~115, no.~32, pp.~8131--8136, 2018.

\bibitem{sen2022equivalence}
S.~Sen, R.~Mandal, and S.~Gangopadhyay, ``Equivalence principle and hbar
  entropy of an atom falling into a quantum corrected black hole,'' {\em
  Physical Review D}, vol.~105, no.~8, p.~085007, 2022.

\bibitem{Wald:1993nt}
R.~M. Wald, ``{Black hole entropy is the Noether charge},'' {\em Phys. Rev. D},
  vol.~48, no.~8, pp.~R3427--R3431, 1993.

\bibitem{MartinMartinez:2011ib}
E.~Mart\'{\i}n-Mart\'{\i}nez, ``Relativistic quantum information: developments
  in quantum information in general relativistic scenarios,'' 2011.

\bibitem{Vidal:2002zz}
G.~Vidal and R.~F. Werner, ``{Computable measure of entanglement},'' {\em Phys.
  Rev. A}, vol.~65, p.~032314, 2002.

\bibitem{Plenio:2005cwa}
M.~B. Plenio, ``{Logarithmic Negativity: A Full Entanglement Monotone That is
  not Convex},'' {\em Phys. Rev. Lett.}, vol.~95, p.~090503, 2005.

\bibitem{Diaz:2023jrf}
B.~D{\'\i}az, D.~Gonz{\'a}lez, M.~J. Hern{\'a}ndez, and J.~D. Vergara,
  ``{Classical analogs of generalized purities, entropies, and logarithmic
  negativity},'' {\em Phys. Rev. A}, vol.~108, no.~1, p.~012411, 2023.

\bibitem{Xu:2024eqg}
B.~Xu, X.~Qi, and J.~Hou, ``{Phase entanglement negativity for bipartite
  fermionic systems},'' {\em Phys. Rev. A}, vol.~110, no.~3, p.~032417, 2024.

\bibitem{Rahaman:2025grm}
A.~Rahaman, ``{Radiative transition of an atom falling into spherically
  symmetric Lorentz violating black hole background},'' 3 2025.

\bibitem{Rahaman:2025mrr}
A.~Rahaman, ``{An atom in front of Lorentz violating Kalb-Ramond black hole
  background},'' 6 2025.

\bibitem{Pantig:2025okn}
R.~C. Pantig and A.~{\"O}vg{\"u}n, ``{Acceleration radiation from
  derivative-coupled atoms falling in modified gravity black holes},'' {\em
  Eur. Phys. J. C}, vol.~85, no.~10, p.~1183, 2025.

\bibitem{Ovgun:2025isv}
A.~{\"O}vg{\"u}n and R.~C. Pantig, ``{HBAR entropy of infalling atoms into a
  GUP-corrected Schwarzschild black hole and equivalence principle},'' {\em
  Phys. Lett. A}, vol.~568, p.~131201, 2026.

\bibitem{Wu:2023vis}
S.-M. Wu, C.-X. Wang, R.-D. Wang, J.-X. Li, X.-L. Huang, and H.-S. Zeng,
  ``{Curvature-enhanced multipartite coherence in the multiverse*},'' {\em
  Chin. Phys. C}, vol.~48, no.~7, p.~075107, 2024.

\bibitem{Bao:2025dzj}
R.~Bao and S.-M. Wu, ``{Spin entanglement and its monogamy relation in
  multi-event horizon spacetime},'' {\em Chin. J. Phys.}, vol.~97,
  pp.~1502--1513, 2025.

\bibitem{Wu:2024nfp}
S.-M. Wu, X.-W. Teng, H.-Y. Wu, J.-X. Li, X.-L. Huang, and R.~Bao, ``{Conserved
  mutual information for discrete and continuous variables in dilaton black
  hole},'' {\em Chin. J. Phys.}, vol.~92, pp.~755--765, 2024.

\bibitem{Liu:2025ctf}
X.~Liu, W.~Liu, and S.-M. Wu, ``{Reversing quantum resource hierarchy:
  non-maximal multipartite entanglement in dilaton spacetime},'' {\em Eur.
  Phys. J. C}, vol.~85, no.~10, p.~1209, 2025.

\bibitem{Li:2024pwo}
S.-H. Li, S.-H. Shang, and S.-M. Wu, ``{Quantum steering for different types of
  Bell-like states in gravitational background},'' {\em Phys. Lett. B},
  vol.~870, p.~139895, 2025.

\bibitem{Du:2024rtc}
M.-M. Du, H.-W. Li, Z.~Tao, S.-T. Shen, X.-J. Yan, X.-J. Y. X.-Y. Li, W.~Zhong,
  Y.-B. Sheng, and L.~Zhou, ``{Basis-independent quantum coherence and its
  distribution under relativistic motion},'' {\em Eur. Phys. J. C}, vol.~84,
  no.~8, p.~838, 2024.

\bibitem{Liu:2025lui}
X.~Liu, W.~Liu, S.-H. Shang, and S.-M. Wu, ``{Does the survival and sudden
  death of quadripartite steering in curved spacetime truly depend on
  multi-directionality?},'' 11 2025.

\bibitem{Perlick:2021aok}
V.~Perlick and O.~Y. Tsupko, ``{Calculating black hole shadows: Review of
  analytical studies},'' {\em Phys. Rept.}, vol.~947, pp.~1--39, 2022.

\bibitem{Cunha:2018acu}
P.~V.~P. Cunha and C.~A.~R. Herdeiro, ``{Shadows and strong gravitational
  lensing: a brief review},'' {\em Gen. Rel. Grav.}, vol.~50, no.~4, p.~42,
  2018.

\bibitem{Liu:2025wwq}
W.~Liu, Y.~Liu, D.~Wu, and Y.-X. Liu, ``{A Universal Framework for
  Horizon-Scale Tests of Gravity with Black Hole Shadows},'' 11 2025.

\bibitem{Liu:2024soc}
W.~Liu, D.~Wu, and J.~Wang, ``{Light rings and shadows of static black holes in
  effective quantum gravity},'' {\em Phys. Lett. B}, vol.~858, p.~139052, 2024.

\bibitem{Liu:2024iec}
W.~Liu, D.~Wu, and J.~Wang, ``{Light rings and shadows of static black holes in
  effective quantum gravity II: A new solution without Cauchy horizons},'' {\em
  Phys. Lett. B}, vol.~868, p.~139742, 2025.

\bibitem{Liu:2024lbi}
W.~Liu, D.~Wu, X.~Fang, J.~Jing, and J.~Wang, ``{Kerr-MOG-(A)dS black hole and
  its shadow in scalar-tensor-vector gravity theory},'' {\em JCAP}, vol.~08,
  p.~035, 2024.

\bibitem{Errehymy:2025ada}
A.~Errehymy, S.~Hansraj, and C.~Hansraj, ``{Black hole Shadows and Null
  Geodesics in Hamaus-Sutter-Wandelt Void Spacetimes with a Quintessential
  Field: Observational Signatures from EHT Data of M87* and Sgr A*},'' {\em
  Astrophys. J.}, vol.~995, no.~2, p.~148, 2025.

\end{thebibliography}

\end{document}